\documentclass{emulateapj}
\usepackage{graphicx,natbib,psfig,textcomp,fontenc}  
\newcommand{\lsim}{\lower 2pt \hbox{$\, \buildrel {\scriptstyle<}\over {\scriptstyle \sim}\,$}}  
\newcommand{\gsim}{\lower 2pt\hbox{$\, \buildrel {\scriptstyle >}\over {\scriptstyle \sim}\,$}}

\shorttitle{}
\shortauthors{}
\begin{document}

\title{HST emission-line images of nearby 3CR radio galaxies:\\ two
  photoionization, accretion and feedback modes}

\author{Ranieri D. Baldi\altaffilmark{1,2}, Javier Rodr\'iguez Zaur\'in\altaffilmark{2,3},
  Marco Chiaberge\altaffilmark{2,4}, \\ Alessandro Capetti\altaffilmark{5}, William B. Sparks\altaffilmark{2}, Ian M. McHardy$^{1}$}
\email{r.baldi@soton.ac.uk}

\altaffiltext{1}{School of Physics and Astronomy, University of Southampton, Southampton, SO17 1BJ, UK}
\altaffiltext{2}{Space Telescope Science Institute, 3700 San Martin Drive,
  Baltimore, Baltimore, MD 21218.}  
\altaffiltext{3}{Department of Physics and Astronomy, University of Sheffield, Sheffield S3 7RH, UK}
\altaffiltext{4}{Johns Hopkins University-Center for Astrophysical Sciences, 3400 N. Charles Street, Baltimore, MD 21218, USA}
\altaffiltext{5}{INAF-Osservatorio Astrofisico
  di Torino, Strada Osservatorio 20, I-10025, Pino Torinese,
  Italy}  

\begin{abstract}

We present {\it HST}/ACS narrow-band images of a low-z sample of 19 3C
radio galaxies to study the H$\alpha$ and [O~III] emissions from the
narrow-line region (NLR).  Based on nuclear emission line ratios, we
divide the sample into High and Low Excitation Galaxies (HEGs and
LEGs). We observe different line morphologies, extended line emission
on kpc scale, large [O~III]/H$\alpha$ scatter across the galaxies, and
a radio-line alignment. In general, HEGs show more prominent emission
line properties than LEGs: larger, more disturbed, more luminous, and
more massive regions of ionized gas with slightly larger covering
factors. We find evidence of correlations between line luminosities
and (radio and X-ray) nuclear luminosities.  All these results point
to a main common origin, the active nucleus, which ionize the
surrounding gas. However, the contribution of additional
photoionization mechanism (jet shocks and star formation) are needed
to account for the different line properties of the two classes.

A relationship between the accretion, photoionization and feedback
modes emerges from this study.  For LEGs (hot-gas accretors), the
synchrotron emission from the jet represents the main source of
ionizing photons. The lack of cold gas and star formation in their
hosts accounts for the moderate ionized-gas masses and sizes. For HEGs
(cold-gas accretors), an ionizing continuum from a standard disk and
shocks from the powerful jets are the main sources of photoionization,
with the contribution from star formation. These components, combined
with the large reservoir of cold/dust gas brought from a recent
merger, account for the properties of their extended emission-line
regions.

\end{abstract} 


\section{Introduction}

Radio galaxies (RGs) are among the most powerful phenomena in the
Universe and represent an important class of objects for two main
reasons. Firstly, they are excellent laboratories to study some of the
most challenging topics in astronomy: the formation and ejection of
relativistic radio jets and its connection to the central black hole
\citep[BH, e.g.][]{Blandford90,Livio99}, the different modes of
accretion onto the BH, the origin and lifetime of the AGN, the AGN
feedback, and how the interstellar medium (ISM) affects the BH
activity
\citep[e.g.][]{Holt03,Hopkins06,Fabian06,heckman14}. Secondly, RGs are
almost invariably associated with the most massive early-type galaxies
(ETGs) in the local universe and likely harbor the most massive BHs
\citep{zirbel96,chiaberge11}. Therefore, RGs can be used as probes of
the formation and evolution of giant ETGs through
cosmic time in relation to their BH activity.

Leaving aside the abundant population of compact radio sources (FR~0s,
\citealt{baldi15a,baldi18}), extended RGs can be classified as FR~I or
FR~II \citep{Fanaroff74}, based on their radio morphology. FR~Is show
bright cores and jets, with the surface brightness decreasing away
from the nuclear region, while FR~IIs with their bright lobes and hot
spots show the opposite trend. Interestingly, FR~Is are usually
located in the center of massive galaxy clusters
\citep[e.g.][]{Owen96} while FR~IIs are generally located in lower
density environments \citep{Hill91,Zirbel97}.

Another classification of RGs is based on the optical emission line
ratios from the nuclear spectrum: high-excitation galaxies and
low-excitation galaxies (HEGs and LEGs)
\citep[e.g.][]{Hine79,laing94,buttiglione10}. HEGs have typically
FR~II radio morphology, while LEGs encompass the two FR classes. The
difference between HEGs and LEGs are thought to be related to a
different mode of accretion. On one hand, HEGs, which are commonly
interpreted as quasar-like AGN, have strong radiation field since they
are powered by a standard accretion disk \citep{Hardcastle09}. Optical
and infrared studies indicate that HEGs show a large amount of cold
gas \citep{baldi08,Dicken09,tadhunter14,westhues16}, which channels
towards the center and fuels the BH, establishing a cold accretion
mode \citep{hardcastle07,mcnamara12,best12,heckman14}.  On the other
hand, LEGs are fuelled by the hot phase of ISM and show evidence for
the presence of a radiatively inefficient accretion disk
\citep{balmaverde08,hardcastle07}. Their nuclei are
synchrotron-dominated since the jet is the main ionizing source
\citep{chiaberge99,hardcastle00,baldi10b,Hardcastle09}.

Ionizing radiation from accreting supermassive BH at
the center of galaxies is believed to illuminate the narrow line
region (NLR) in the host galaxies of AGN, creating emission-line regions
which can extend up to kilo-parsec scales (e.g \citealt{bennert02,congiu17,sun18}).

Although the efficiency of the AGN feedback is still matter of
discussion in local galaxies (see \citealt{shangguan18}), there are
growing observational evidence of the AGN feedback on the ISM in
radio AGN
(e.g. \citealt{Tremblay09,fabian12,couto13,harrison15,roche16,couto17,harrison18}). In
the so-called {\it quasar-mode feedback} scenario
(e.g. \citealt{silk98,fabian99}), the central engine efficiently
ionizes the innermost and an intermediate region of the NLR (within
some kpc scale, e.g. \citealt{gomez17}) which show a density and
ionization stratification
\citep{balmaverde16,adhikari16}. Furthermore, with a radial decrease
of efficiency, its radiation field can ionize the medium at larger
distance from the nucleus up to some kpc
\citep[e.g.][]{Robinson87,Baum89a}. Such extended emission has a rich
spectrum which is often characterized by large
[O~III]$\lambda$5007/H$\beta$ and HII$\lambda$4686/H$\beta$ ratios
and, in some cases, includes very high ionization species such as
[Ca~V], [Fe~VIII] or [Fe~X]
\citep[e.g.][]{Fosbury82,Tadhunter87,erkens97}. The observed line
ratios are usually explained in terms of photoionization by a central
source on clouds formed by a combination of optically thin and
optically thick materials \citep{Binette96,Robinson00}.  Conversely,
the so-called {\it radio-mode feedback}
(e.g. \citealt{croton06,bower06}) is released in a kinetic form via
radio jets, which deposit energy into the ISM and ionize the gas
\citep{capetti05b}. The majority of the local RGs are currently in a
radiatively inefficient accretion regime, where kinetic feedback
dominates over radiative feedback \citep{ishibashi14}. The two
feedback modes can coexist or delayed in time in RGs, regulated by the
different accretion modes and duty cycles \citep{okamoto08,birzan12}.

Other sources, which produce high-energy photons, could have a
secondary role in the photoionization of the gas in RGs: fast
radiative shocks \citep[e.g.][]{Evans85,Dopita95,Clark98,Groves04},
hot stars, \citep{allen08}, and relativistic particles associated with
jets \citep{lea78}.

Assuming photoionization by a central source as the dominant
mechanism, the emission line luminosities from the ionized gas at
large scales may be used to derive the properties of the central
engine, such as the shape of the photoionizing continuum and its
power \citep[][]{Morganti91,Tadhunter94}. For example, the well
established correlations between the optical emission lines and the
178-MHz radio power \citep[e.g.][]{Rawlings91,Tadhunter98}, suggests a
direct link between the AGN continuum and the mechanism responsible of
the radio-jet production. A relation between the radio and ionized
gas, as stellar content and emission lines, has been addressed as the
'alignment effect', where the optical and infrared light is aligned
along radio-jet axes (e.g. \citealt{mcharthy87,chambers87}). It has
been explained as the effect of either the dust scattering of light
from a hidden quasar at the nucleus or triggering of star formation by
the synchrotron jet as it propagates outward from the nucleus.
Overall, although emission lines provide a simple tool to study the
NLR, the details of the mechanisms involved in the production of such
ionized gas are yet to be fully understood.

In this context, the {\it Hubble Space Telescope} ({\it HST}) offers a
unique opportunity to study the spatial properties of the line
emitting gas with unprecedented resolution. In this paper we present
narrow-band optical {\it HST}/ACS observations sampling the
H$\alpha$+[N~II] and [O~III]$\lambda$5007 emission for 19 3C RGs with
z $<$0.3. \cite{Tremblay09} presented the data-set and a preliminary
analysis of the images. Here we have used the most up-to-date version
of the {\it HST}/ACS reduction pipeline that allows us to improve the
quality of the data-set, which is again presented and analyzed in
detail. The HST/ACS high-resolution data-sets allow us to study the
properties of the emission line region and their variation across the
NLR in the perspective of the HEG-LEG scheme. In addition, we combine
our images with existing optical ground-based spectroscopic data
\citep{buttiglione09,buttiglione10,Buttiglione11} (hereafter B09+) and
      {\it Chandra} X-ray observations \citep[][Torresi et al. in
        prep]{massaro10,massaro12,Balmaverde12} to estimate a number
      of physical parameters, such as the number of ionizing photons,
      covering factors, and mass of ionized gas, with the aim of
      better understanding the nature of the ionizing source and the
      properties of the ionized gas itself.

\tabletypesize{\scriptsize}
\setlength{\tabcolsep}{0.25pc}
\begin{deluxetable}{cccccccc}
\tablecolumns{8}
\tabletypesize{\footnotesize}
\tablecaption{ \label{detect} Observing log}
\startdata
\tableline
\addtolength{\columnsep}{-1cm}
Name & z &  RA        & DEC       & Filter & Line/Cont & $\lambda_{c}$ & $t_{exp}$ \\
 3C    &   &  (J2000.0) & (J2000.0) &        &           &    (\AA)      &   (s) \\
(1)  &(2)&    (3)     & (4)       &  (5)   &   (6)     &     (7)       & (8) \\
\tableline
33.0  & 0.0597 & 01 08 52.8 & +13 20 14  & FR647M & Cont & 5828 & 60 \\
      &        &            &            & FR716N & H$\alpha$ & 6951 & 400  \\
      &        &            &            & FR551N & [O~III] & 5303 & 500\\
40.0  & 0.0180 & 01 25 59.9 & -01 20 33  & FR647M & Cont & 5599 &  60 \\ 
      &        &            &            & FR656N & H$\alpha$ & 6674 & 400 \\
      &        &            &            & FR505N & [O~III] & 5092 & 500\\
78.0  & 0.0286 & 03 08 26.2 & +04 06 39  & FR647M & Cont & 5575 & 60 \\
      &        &            &            & FR656N & H$\alpha$ & 6746 & 400 \\
      &        &            &            & FR505N & [O~III] & 5147 & 500 \\
93.1  & 0.2430 & 03 48 46.9 & +33 53 15  & FR647M & Cont & 6837 & 60 \\
      &        &            &            & FR853N & H$\alpha$ & 8164 & 400 \\
      &        &            &            & FR601N & [O~III] & 6228 & 500\\
129.0 & 0.0208 & 04 49 09.1 & +45 00 39  & FR647M & Cont & 5620 & 60 \\
      &        &            &            & FR656N & H$\alpha$ & 6700 & 400 \\
      &        &            &            & FR505N & [O~III] & 5112 & 500\\
132.0 & 0.2140 & 04 49 09.1 & +45 00 39  & FR647M & Cont & 6674 & 60\\
      &        &            &            & FR782N & H$\alpha$ & 7966 & 400 \\
      &        &            &            & FR601N & [O~III] & 6078 & 500\\
136.1 & 0.0640 & 04 56 43.0 & +22 49 22  & FR647M & Cont & 5856 & 60 \\
      &        &            &            & FR716N & H$\alpha$ & 6983 & 400 \\
      &        &            &            & FR551N & [O~III] & 5328 & 500\\
180.0 & 0.2200 & 07 27 04.5 & -02 04 42  & FR647M & Cont & 6707 & 60 \\
      &        &            &            & FR782N & H$\alpha$ & 8005 & 400\\
      &        &            &            & FR601N & [O~III] & 6108 & 500\\
196.1 & 0.1980 & 08 15 27.8 & -03 08 27  & FR647M & Cont & 6587 & 60 \\
      &        &            &            & FR782N & H$\alpha$ & 7861 & 400 \\
      &        &            &            & FR601N & [O~III] & 5998 & 500\\
197.1 & 0.1280 & 08 21 33.6 & +47 02 37  & FR647M & Cont & 6217 & 60 \\
      &        &            &            & FR716N & H$\alpha$ & 7414 & 400\\
      &        &            &            & FR551N & [O~III] & 5657 & 500\\
219.0 & 0.1744 & 09 21 08.6 & +45 38 57  & FR647M & Cont & 6456 & 60 \\
      &        &            &            & FR782N & H$\alpha$ & 7704 & 400\\
      &        &            &            & FR601N & [O~III] & 5879 & 500 \\
227.0 & 0.0858 & 09 47 45.1 & +07 25 20  & FR647M & Cont & 5976 & 60 \\
      &        &            &            & FR716N & H$\alpha$ & 7127 & 400\\
      &        &            &            & FR551N & [O~III] & 5438 & 500\\
234.0 & 0.1849 & 10 01 49.5 & +28 47 09  & FR647M & Cont & 6510 & 60 \\
      &        &            &            & FR782N & H$\alpha$ & 7770 & 400\\
      &        &            &            & FR601N & [O~III] & 5928 & 500\\
270.0 & 0.0075 & 12 19 23.2 & +05 49 31  & FR647M & Cont & 5546 & 60\\
      &        &            &            & FR656N & H$\alpha$ & 6611 & 400\\
      &        &            &            & FR505N & [O~III] & 5044 & 500\\
285.0 & 0.0794 & 13 21 17.9 & +42 35 15  & FR647M & Cont & 5938 & 60\\
      &        &            &            & FR716N & H$\alpha$ & 7081 & 400 \\
      &        &            &            & FR551N & [O~III] & 5403 & 500\\
314.1 & 0.1197 & 15 10 22.5 & +70 45 52  & FR647M & Cont & 6156 & 60\\
      &        &            &            & FR716N & H$\alpha$ & 7342 & 400\\
      &        &            &            & FR551N & [O~III] & 5602 & 500\\
319.0 & 0.1920 & 15 24 05.6 & +54 28 18  & FR647M & Cont & 6554 & 60\\
      &        &            &            & FR782N & H$\alpha$ & 7822 & 400\\
      &        &            &            & FR601N & [O~III] & 5968 & 500\\
388.0 & 0.0917 & 18 44 02.4 & +45 33 30  & FR647M & Cont & 6007 & 60\\
      &        &            &            & FR716N & H$\alpha$ & 7164 & 400\\
      &        &            &            & FR551N & [O~III] & 5466 & 500\\
390.3 & 0.0561 & 18 42 09.0 & +79 46 17  & FR647M & Cont & 5812 & 60\\
      &        &            &            & FR716N & H$\alpha$ & 6931 & 400\\
      &        &            &            & FR551N & [O~III] & 5288 & 500\\
\enddata
\tablecomments{Log of the {\it HST}/ACS WFC observations (Program
  10882, PI: W.B. Sparks). Col (1): 3C source name. Col (2): redshift. Col
  (3)-(4): right ascension and declination.  Col (5)-(6): ramp filters
  used for the continuum, H$\alpha$ and [O~III] observations,
  respectively. Col (7) central wavelength observed corresponding to
  the ramp filter configuration. Col (8): total exposure time.}
\end{deluxetable}

We organize this paper as follows. Section 2 presents a description of
the sample and the corresponding observations, while the reduction
process and calibration of the data are presented in Section 3.
Section 4 shows the catalogue of H$\alpha$+[N~II] and [O~III] images
and the corresponding flux measurements. In addition, we show
[O~III]/H$\alpha$+[N~II] emission line ratio images. Section 4
presents the results of measuring the physical parameters mentioned
before. In Section 5 we discuss all the previous results in light of
the two AGN classes (LEGs and HEGs). Summary and potential future
observations are presented in Section 6. Throughout this paper we use
H$_{0}$ = 71 km s$^{-1}$, $\Omega_{\rm M}$ = 0.27 and
$\Omega_{\Lambda}$ = 0.73.

\begin{table}
\begin{center}
\caption{Multi-band properties}
\begin{tabular}{lcccccc}
\tableline\tableline
Name &  FR & optical & log L$_{178}$ & log L$_{core}$  & log L$_{X}$ & $\Gamma$ \\
 3C  &  class & class   & erg s$^{-1}$Hz$^{-1}$ & erg s$^{-1}$Hz$^{-1}$ & erg s$^{-1}$ & \\
(1)  &  (2)  & (3) & (4) & (5) & (6) & (7) \\
\tableline
33   & II    & HEG &33.65 & 30.27 & 43.74   &  1.70(fix)  \\
40   & I     & LEG &32.29 & 30.66 & 41.06   &  1.77   \\
78   & I     & LEG &32.51 & 31.25 & 42.30   &  2.20   \\
93.1 & II    & HEG &34.24 & $-$   & 43.25   &  1.75   \\
129  & I     & LEG &32.65 & 28.92 & 40.09   &  2.32   \\
132  & II    & LEG & 34.26& 31.40 & $<$41.99$^{a}$&  2.35   \\
136.1& II    & HEG &33.13 & 29.16 & $<$42.91&  1.70(fix) \\
180  & II    & HEG &34.32 & $-$   & 43.84   &  1.70(fix)  \\
196.1& II    & LEG &34.31 & 31.65 &  $-$    &  $-$   \\
197.1& II    & BLO &33.55 & 30.30 & 43.62   &  1.30  \\
219  & II    & BLO &34.53 & 31.54 & 44.14$^{b}$   &  1.58  \\
227  & II    & BLO &33.74 & 30.48 & 43.27$^{c}$   &  1.5(fix)  \\ 
234  & II    & HEG &34.47 & 31.88 & 44.05   &  1.5    \\
270  & I     & LEG &32.75 & 29.50 & 40.70$^{d}$   &  0.8   \\
285  & II    & HEG &33.23 & 29.93 & 43.36   &  2.0(fix)  \\
314.1& double& ELEG&33.59 & 29.36 &  $-$    &  $-$      \\
319  &  II   & LEG &34.20 & 29.93 & $<$42.71&  1.7(fix)   \\
388  & II    & LEG &33.70&  31.04 & 41.87$^{a}$   &  2.28  \\
390.3& II    & BLO &33.54 & 31.38 & 44.38   &  1.83  \\
\tableline
\end{tabular}
\label{multiband}
\tablecomments{Col (1): 3C source name.  Col (2): morphological FR
  type. Col (3): spectroscopic class from \cite{buttiglione10} - LEG =
  low excitation galaxy, HEG = high excitation galaxy, ELEG =
  extremely low excitation galaxy, BLO = broad line object. Col (4):
  Log of the radio power at 178 MHz. Col (5): Log of the radio core
  power at 5 GHz. References can be found in \cite{buttiglione10} and
  \cite{baldi10b} Col (6): Log X-ray (2-10 keV) unabsorbed nuclear
  luminosities from {\it Chandra} data from Torresi et al. (in
  prep). The exceptions are: \cite{Hardcastle09} marked by $^{a}$,
  \cite{shi05} marked by $^{b}$, \cite{kataoka11} marked by $^{c}$,
  \cite{balmaverde06a} marked by $^{d}$. Typical error on the X-ray
  luminosities is 0.2 dex. Col (7): X-ray photon index $\Gamma$
  (references from Col 6.).}
\end{center}
\end{table}

\section{Sample selection and observations}

The sample selection and observations are described in detail in
\cite{Tremblay09}. To summarize, we present a sample of 19 3CR radio
sources from the low-z (z $<$ 0.3) extra-galactic radio-source Revised
Third Cambridge Catalogue
\citep[3CR,][]{Bennett62a,Bennett62b,Spinrad85}. The sample comes from
the original HST proposal for 98 3CR sources that was awarded for
observations prior to the ACS side 2 electronics failure in January
2007. Although only 19 sources were observed (Table~\ref{detect}),
these objects represent an unbiased sample in selection due to the
nature of the {\it HST} snapshot mode (to fill gaps in the {\it HST}
schedule, sources are randomly picked from the target list). Moreover,
the observed 19 objects span almost the whole redshift range (0.0075
$<$ z $<$ 0.224) of the original sample and show a wide variety of
emission line ratios, radio power or spectroscopic properties. With
regards to their radio morphologies, the sample consists of 4 FR~Is
and 14 FR~IIs, taking the double-lobed 3C~314.1 apart
(Table~\ref{multiband}).  In terms of their optical, spectroscopic
classification, 8 of the objects in our sample are classified as LEGs,
10 are HEGs, of which four have broad lines, named Broad Line
Objects (BLOs). Of particular interest is 3C~314.1, a RG whose
spectral classification is Extremely Low-Excitation Galaxy (ELEG,
\citealt{capetti11}) interpreted as relic RG
\citep{capetti13}. Table~\ref{multiband} summarizes the radio and
spectroscopic classification of the sample. This table also
  collects the total radio luminosities at 178 MHz, the radio core
  powers (at 5 GHz) and the X-ray luminosities for the sample. The
  uncertainties on the radio luminosities are negligible with respect
  to those of the X-ray luminosities, which amount to $\sim$0.2 dex.

\begin{table*}
\begin{center}
\caption{HST optical properties}
\begin{tabular}{ccccccccc}
\tableline\tableline
Name & H$\alpha$+[N~II] flux & [O~III] flux& L$_{\rm H\alpha +  [N~II]}$ & L$_{\rm [O~III]}$ & P$_{\rm [N~II]}$ & P$_{\rm [O~III]\lambda\rm 4959}$ &  BLR  & R$_{\rm flux}$ \\
 3C  &  10$^{-14}$ erg s$^{-1}$ cm$^{-2}$ & 10$^{-14}$ erg s$^{-1}$ cm$^{-2}$ & erg s$^{-1}$ & erg s$^{-1}$ & \% & \% & \% & kpc \\
(1)  & (2) & (3) & (4) & (5) & (6) & (7) & (8) & (9) \\
\tableline
33.0 & 8.61$\pm$0.70 & 12.86$\pm$0.94 & 41.85 & 42.03 & 50 & 10 & -    & 0.95 \\       
40.0 & 1.50$\pm$0.14 & 0.20$\pm$0.03  & $<$40.02  & $<$39.16 & -  & - & - & 0.25\\      
78.0 & 5.05$\pm$0.80 & 1.02$\pm$0.11 & 40.96 &  40.27 & 81 & -& - & 0.14\\ 
93.1  &3.02$\pm$0.31 & 2.61$\pm$0.08 & 42.72 & 42.66 & 65 & 12      & - & 1.03\\   
129.0 &1.05$\pm$0.21 & -    & 40.00 & - & -  & - &             & 0.03 \\  
132.0 &0.22$\pm$0.05 &   -  & 41.46 &  -& -  & - &             & 0.59\\ 
136.1 &6.36$\pm$0.75 & 2.68$\pm$0.21 & 41.78 & 41.41 & 47 & -       & - & 0.55 \\ 
180.0 &2.83$\pm$0.61 & 7.02$\pm$1.20 & 42.60 & 42.99 & 50 & 13      & - & 7.00 \\ 
196.1 &1.34$\pm$0.26 & 0.35$\pm$0.06 & 42.17 & 41.58 & 62 & 10      & - & 3.39 \\ 
197.1 &0.91$\pm$0.02 & 0.32$\pm$0.06 & 41.58 & 41.13 & 25 & -       & 58 & 0.27 \\ 
219.0 &2.88$\pm$0.42 & 1.56$\pm$0.25 & 42.38 & 42.11 & 30 & -       & 52 & 0.40\\ 
227.0 &22.99$\pm$1.82 & 5.67$\pm$0.68 & 42.61 & 42.01 &  1 & 25 & 95 & - \\ 
234.0 &15.03$\pm$1.41 & 25.55$\pm$2.60 & 43.15 & 43.38 & 22 & -      & -  & 0.76 \\ 
270.0 &3.90$\pm$0.73 & 0.39$\pm$0.08 & $<$39.68 & $<$38.67 & -  & - & -  & 0.07\\
285.0 &1.07$\pm$0.25 & 0.91$\pm$0.20 & 41.21 & 41.14 & 45 & -       & - & 1.77 \\ 
314.1 &0.10$\pm$0.02 & 0.03$\pm$0.01 & 40.54 & 40.02 & 53 & - & - & 0.36 \\ 
319.0 &0.14$\pm$0.03 & -  & 41.16 & - &  - & - & -& 0.72\\ 
388.0 &1.78$\pm$0.25 & 0.18$\pm$0.04 & 41.56 & 40.57 & 76 & -       & - & 0.32 \\ 
390.3 &64.96$\pm$5.52 & 24.18$\pm$2.1 &  42.67 & 42.25 &  5 & 30  & 88 & - \\ 
\tableline
\end{tabular}
\label{detect2}
\tablecomments{Col (1): 3C source name. Col (2): total
  H$\alpha$+[N~II] flux. Col (3): total [O~III] flux. Col (4):
  total H$\alpha$+[N~II] luminosity. Col (5): total [O~III]
  luminosity. Col (6): estimated
  [N~II]$\lambda$$\lambda$6549,6583\AA~percentage contribution to the
  flux within the wavelength range of the filter. Col (7): estimated
  [O~III]$\lambda$4959\AA~contribution to the flux within the
  wavelength range of the filter. Col (8): percentage of the emission
  that corresponds to the broad H$\alpha$ component for those sources
  spectroscopically classified as BLOs.  Col (9): ``flux weighted
  radius'' (see text for details).
}
\end{center}
\end{table*}

Our {\it HST} data-set consists of images taken with the ACS Wide
Field Channel (WFC). Linear ramp filters, with typical
$\lambda$ range of 4500-7500 \AA\, were used to image the
[O~III]$\lambda$5007 and the H$\alpha$+[N~II] (hereafter H$\alpha$ for
simplicity) emission from the galaxies in our sample, adjusting the
central wavelength according to the redshifts of the objects. The
details of the observations for the 19 galaxies of our sample
are presented in Table~\ref{detect}.

\section{Data reduction and calibration}

The data were reduced using the standard data reduction pipeline
procedures, which employ two packages: the CALACS package, that
includes dark subtraction, bias subtraction and flat-field corrections
and produces calibrated images, and the MULTIDRIZZLE package, which
corrects for distortion and performs cosmic ray rejection.

In order to reject cosmic rays, hot pixels and other artifacts, two
exposures of 250 and 200 sec were taken for the [O~III] and the
H$\alpha$ emission lines, respectively. In addition, with the aim of
generating ``pure'' emission line images, the continuum emission
between the two lines was sampled using 60 seconds single exposures of
the medium ramp filter FR647M centered at various wavelengths
(5370-7570 \AA), depending on the wavelength of the target.  For these
images, the cosmic rays were rejected ``manually'' using the routines
{\rm IMEDIT} in IRAF and/or {\rm CLEAN} within the {\rm STARLINK}
package {\rm FIGARO}.

\begin{figure*}
\hspace{1.75cm}\includegraphics[scale=0.8]{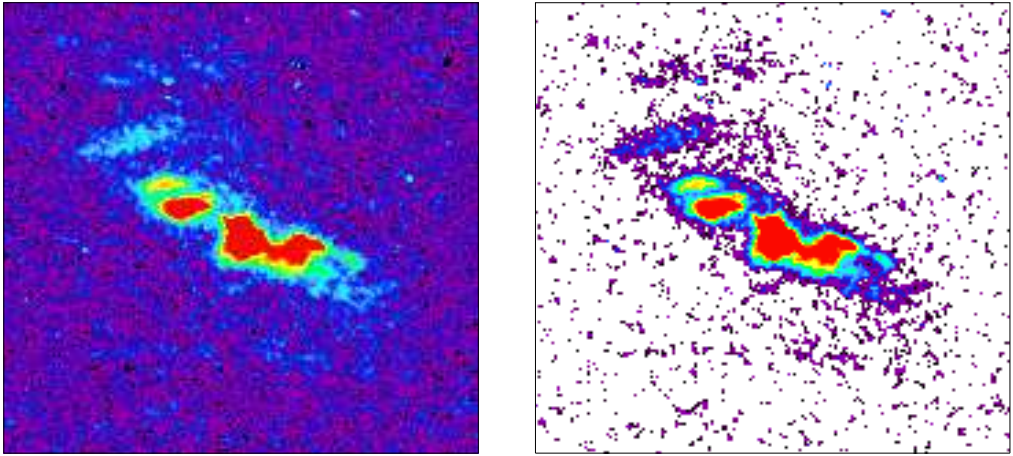}
\caption{An example of the ``source region'' selection process for the
  galaxy 3C~33. On the left the raw image and on the right the final
  continuum-subtracted image. In the color print version, redder
  color corresponds to higher emission line brightness.}
\label{3c33}
\end{figure*}

The background was subtracted from the images using the average value
(in electrons s$^{-1}$) of a region located at the outskirts of the
galaxies, avoiding the central regions. 

To convert into physical units the background-subtracted images, it is
possible to use, a priori, the PHOTFLAM keyword. PHOTFLAM is the
inverse sensitivity (erg cm$^{-2}$ sec$^{-1}$ \AA$^{-1}$) and
represents the flux of a source with constant F$_{\lambda}$ which
produces a count rate of 1 electron per second. Given that the ACS
drizzled images have units of electrons s$^{-1}$, flux units of erg
cm$^{-2}$ sec$^{-1}$ \AA$^{-1}$ are obtained just by multiplying the
images by the PHOTFLAM value. However, while this is adequate when the
flux is approximately constant through the bandpass, it is certainly
not adequate when the spectrum of the source shows a significant slope
or strong emission lines within the bandwidth, which is the case for
most of our observations.   Therefore, in
order to adequately calibrate the 3C images in flux, we first examined
the spatial shape of the corresponding long-slit, optical spectra
available for {\it all} the galaxies in our sample (B09+). 

The PHOTFLAM parameter was finally used for all continuum images and
for those emission-line images for which the corresponding spectra
show weak emission lines in the nuclear region. Conversely, for those
galaxies with evidence of significant [O~III] and H$\alpha$ emission,
it is more convenient to use the Equivalent Monochromatic Flux (EMFLX)
of the filter. This parameter is obtained using the routine BANDPAR
within SYNPHOT package in IRAF, and is calculated as follows:

\begin{equation} 
EMFLX = URESP \times \frac{\int P_{\lambda} d\lambda}{P_{\lambda cen}}
\end{equation}

where URESP is the response of the filter (i.e. the flux that produces
1 count per second in the passband), P$_{\lambda}$ is the
dimensionless passband throughput as a function of wavelength and
P$_{\lambda cen}$ is a passband throughput at the central wavelength
of the filter. The data calibration performed by
  \cite{Tremblay09} did not include this spectral correction across
  the bandwidth for the sources.

The next step during the reduction process is to align the continuum
images with the corresponding emission line images. We used the
routines GEOMAP and GEOTRAN in IRAF to geometrically align the images
in those cases for which 3 or more foreground stars (less than 10)
were present in the field of view. For the remaining cases we used the
routines IMSHIFT and ROTATE in IRAF to shift and rotate the images
based on features observed in the galaxy and/or present in the frame.
To check the accuracy of the alignment we used the routine CENTER in
IRAF to calculate the centroid of the stars in the aligned images and
the position of some of the mentioned features. We estimate that the
alignment accuracy is usually better than 0.5 pixels.

Once the images were aligned, the continuum emission was subtracted
from the emission line images in order to produce ``pure'' [O~III] and
H$\alpha$ images. Finally, we selected a 2$\sigma$ lower cut as a
threshold to distinguish galaxy from background, where $\sigma$ was
the root mean square (rms) of the flux in a region in the field of
view far from galaxy emission. The 2$\sigma$ lower cut procedure was
selected based on visual inspection of contour plots onto the
continuum-subtracted emission line images. All pixels with flux values
below the limit threshold were set to 0. The emission line flux from
the galaxy was then measured by integrating within the limiting
contour level, i.e. the contour that separates galaxy from
background. Figure~\ref{3c33} shows an example of the process for the
case of 3C~33.

\subsection{Flux calibration accuracy}

To check the flux calibration accuracy we compared the flux
measurements from our images with those obtained by B09+ spectra, by
convolving the spectra with the spectral response of the {\it HST} ACS
ramp filters on the same integration area. Since B09+ spectra were
extracted using a 2$\arcsec$ $\times$ 2$\arcsec$ box, we first
analyzed the objects for which the entire emission in our images was
confined in a similar area to avoid uncertainties due to the presence
of extended emission. We then applied the same procedure to all
galaxies in our sample, centering a 2$\arcsec \times 2\arcsec$ box on
the peak of the continuum emission in our ACS images. We found that
the flux values derived using our HST images were, in general,
consistent within a factor of 2 with respect to those from the
long-slit spectroscopy. The exceptions are 3C~40, 3C~78 and 3C~270. In
the case of 3C~78, the continuum image has a cosmic ray coinciding
with the location of the nucleus. Since it is not possible to entirely
correct for this effect during the reduction process, the continuum
subtraction is not highly accurate for this source. For 3C~40 and
3C~270, the difference is due to fact that HST images are probably
dominated by the continuum (see Sect. 4.1).

The images are not corrected for the [N~II]$\lambda\lambda$6549,6583
or the [O~III]$\lambda$4959 contribution to the flux that falls within
the bandpass of the ramp filters for the H$\alpha$ and the
[O~III]$\lambda$5007, respectively.  To give an idea of such
contributions we have used the B09+ optical spectra along with the
STARLINK package DIPSO to model the emission line profiles with one or
two Gaussians per emission line. The values are presented in
Table~\ref{detect2}. Furthermore, the spectra of BLOs show impressive
broad H$\alpha$ and H$\beta$ components which strongly contribute to
the emission of the narrow H$\alpha$ and [O~III] observations,
respectively. We estimated the percentage contribution of H$\alpha$ and
H$\beta$ broad component ($\sim$50-95 per cent) to correct the total
emission-line luminosities for BLOs in Table~\ref{detect2}, using the
spectra from B09+.

\section{Results}

\subsection{The H$\alpha$ and [O~III] images }

\begin{figure*}
\includegraphics{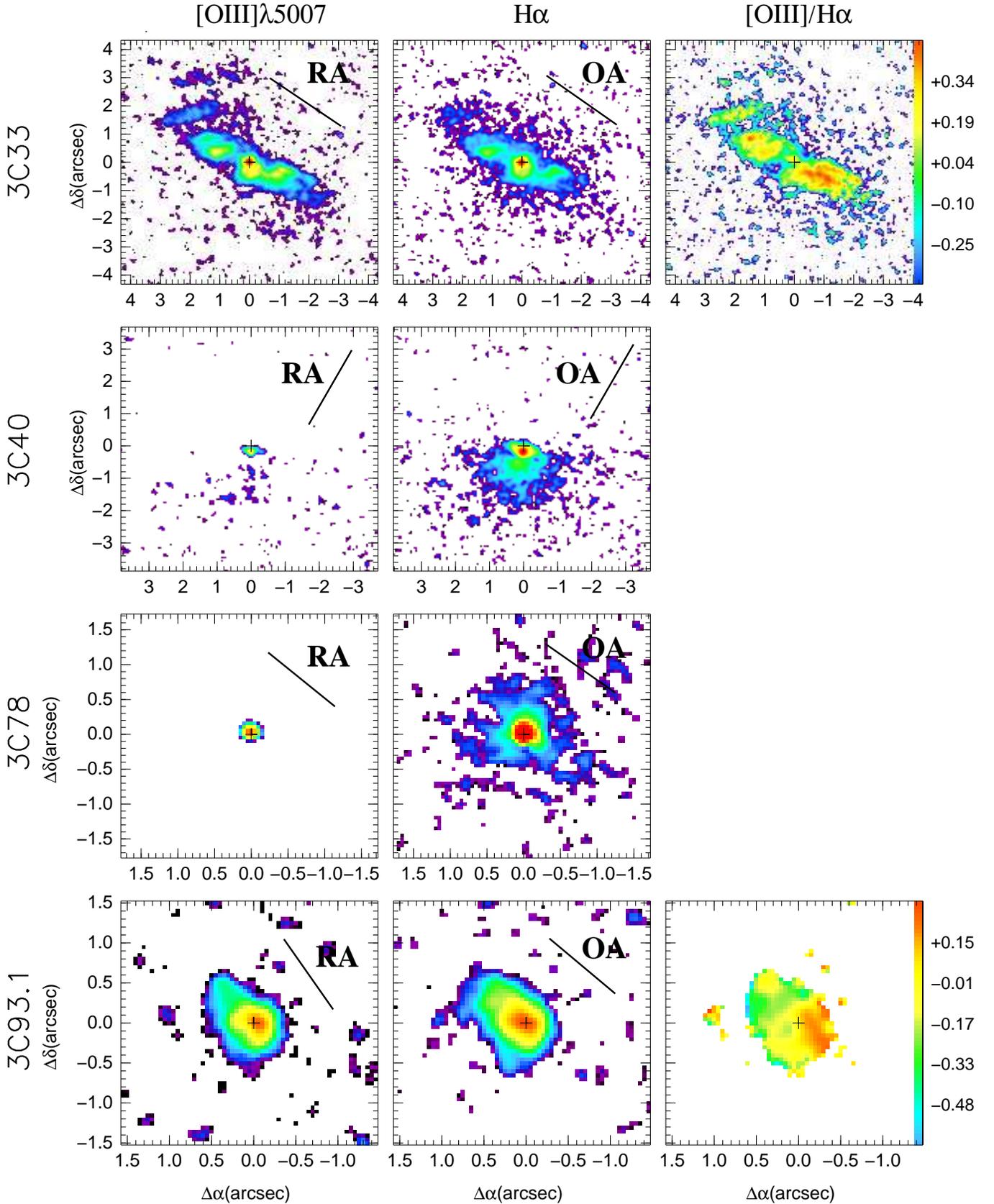}
\caption{{\it Left and middle columns}: the H$\alpha$ and
  [O~III]$\lambda$5007 continuum-subtracted images of the
  sources. {\it Right column}: images showing the spatial distribution
  of the logarithm of the [O~III]/H$\alpha$ flux ratio, for those
  galaxies with extended emission and/or sufficient S/N to generate
  reliable line ratio images. The plus sign indicates the center of
  the galaxy defined as the maximum of the continuum emission in the
  corresponding continuum observation. The black-solid lines in the
  [O~III] and H$\alpha$ images indicate the orientation of the radio
  (RA) and optical axis (OA) respectively (see text for details).  For
  those objects with no clear radio and/or optical axis no lines are
  shown in the Figure. All images are represented in logarithmic
  scale. For all images North is top, East is left.}
\label{panel}
\end{figure*}

\addtocounter{figure}{-1}

\begin{figure*}
\includegraphics{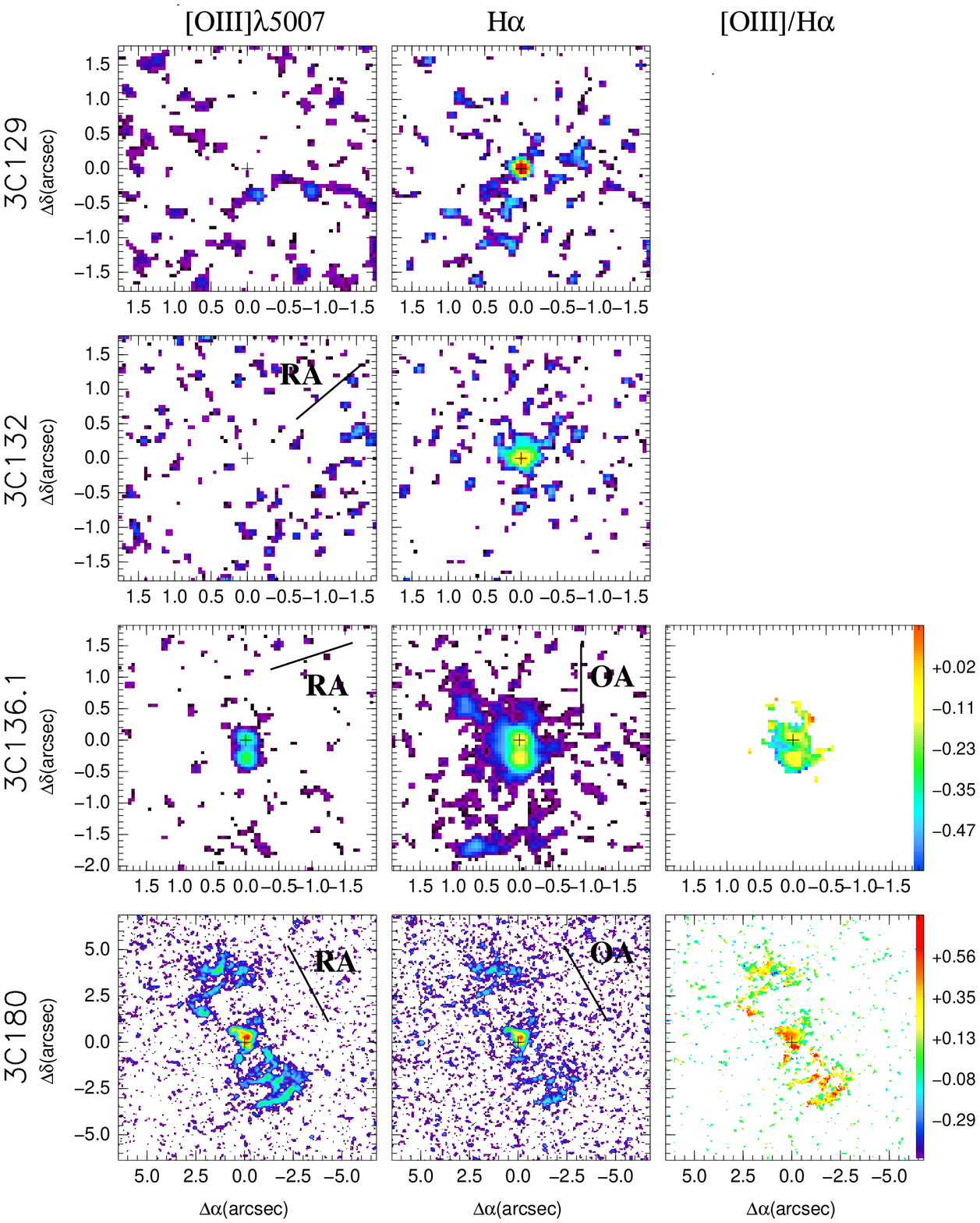}
\caption{Continued}
\end{figure*}

\addtocounter{figure}{-1}

\begin{figure*}
\includegraphics{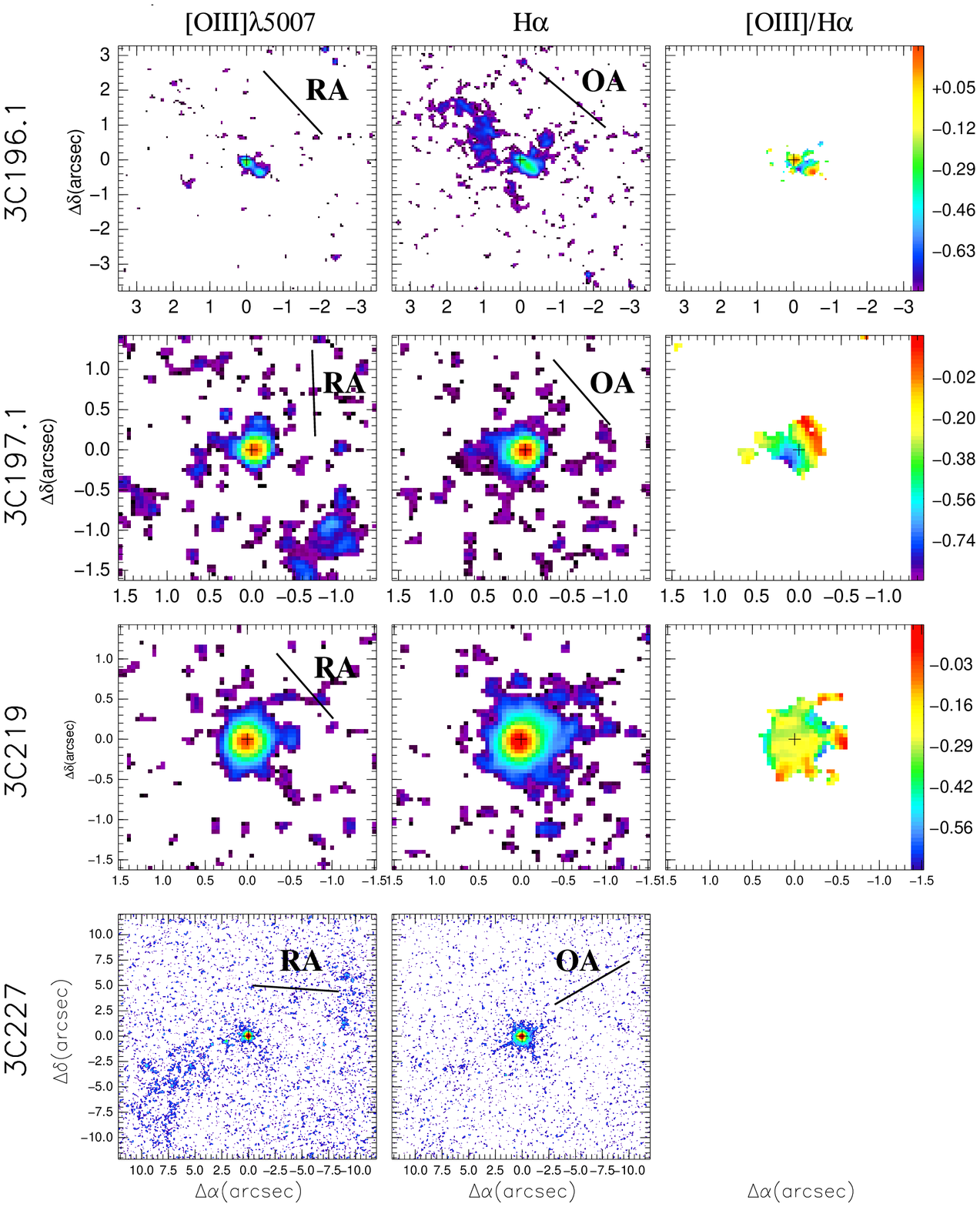}
\caption{Continued}
\end{figure*}

\addtocounter{figure}{-1}

\begin{figure*}
\includegraphics{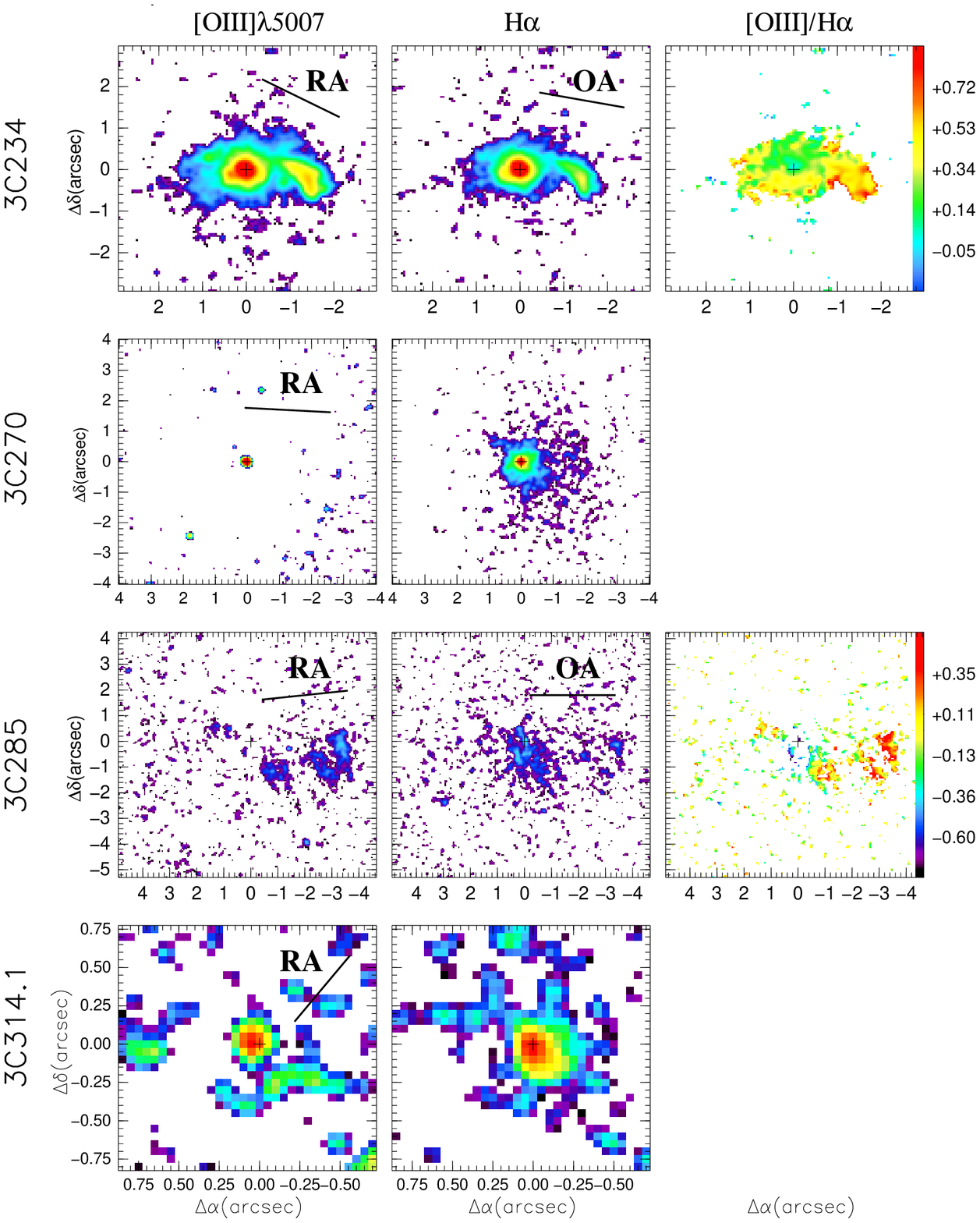}
\caption{Continued}
\end{figure*}

\addtocounter{figure}{-1}

\begin{figure*}
\includegraphics{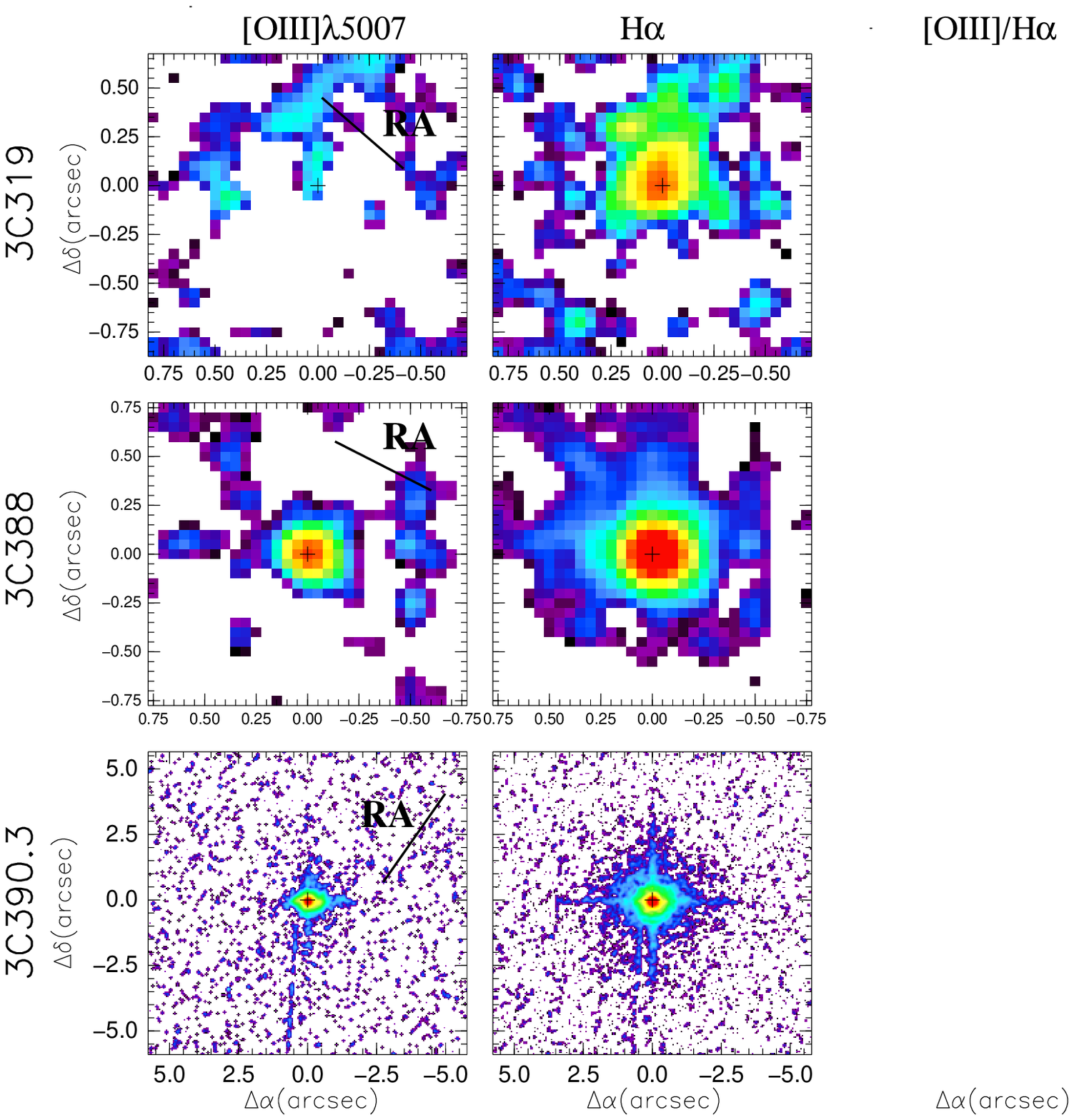}
\caption{Continued}
\end{figure*}

Figure~\ref{panel} (first and second columns) is a collection of the
[O~III] and H$\alpha$ continuum-subtracted images of all
galaxies in our sample. Table~\ref{detect2} presents the integrated
[O~III] and H$\alpha$ flux values measured from those maps, which
  are corrected for H$\alpha$ and H$\beta$ broad line contribution and
  not for the [O~III]$\lambda$4959 and [N~II] narrow lines. In
addition, the values were corrected for Galactic reddening using the
\cite{Howarth83} extension to optical wavelengths of the
\cite{Seaton79} reddening law, along with the E(B - V) values derived
from the far-IR based maps of extinction by \cite{Schlegel98}. 
  The mean estimated uncertainty of the fluxes is 15\%, ranging
  between 8 and 30\%.

The up-to-date version of the {\it HST}/ACS reduction pipeline, the
correction for the spectral slope in the filter bandwidth and the
correction for broad and narrow emission lines result in measured
emission-line fluxes which significantly differ from the values
derived by \cite{Tremblay09} by 1--2 orders of magnitude. In addition,
the line luminosities derived from our images generally agree with the
values extracted from the long-slit spectra (B09+) for the sources
with compact line emission. For the sources with extended line
emission, our luminosities are higher than those derived from the
nuclear spectra within a factor 10.

Ionized gas emission, either [O~III] or H$\alpha$ is detected for the
majority of objects in our sample. However, there are some cases that
deserve more attention. On one hand, the optical spectra of 3C~40 and
3C~270 show weak or no evidence of ionized gas emission
(B09+). Conversely, both our H$\alpha$ and [O~III] images clearly
reveal enhanced emission concentrated toward the nuclear region of the
sources. Therefore, it is likely that their HST images are
dominated by continuum and the emission lines are diluted by such a
component. In this case, the emission-line fluxes, derived from the HST
images, would be upper limits (Table 3). On the other hand, the HST
images of 3C~129, 3C~132 and 3C~319 show no evidence for [O~III]
emission, but a faint compact emission in the H$\alpha$ images
emerges. This evidence is consistent with their optical spectra from
B09+ which show detected H$\alpha$ and weak or non-detected [O~III].

Considering the whole sample, we found that the H$\alpha$ and [O~III]
luminosities range from 9.3$\times$10$^{39}$ to 1.4$\times$10$^{43}$
erg s$^{-1}$ and 4.6$\times$10$^{38}$ to 2.4$\times$10$^{43}$ erg
s$^{-1}$. HEGs show larger emission-line luminosities than LEGs, on
average, by a factor $\sim$26 and and $\sim$140, respectively for
  H$\alpha$ and [O~III].

\subsection{Morphology and size of the ionized gas emission}

The emission-line images in Figure~\ref{panel} show a wide range of
different morphologies. On one hand, HEGs show extended and disturbed
morphologies on the scale of some kpc.  Spectacular are the cases of
3C~33 and 3C~180. Their extended emission-line regions (ELRs) show
tails, bridges, shells, irregular features and amorphous halos. The
broad-lined HEG, BLOs, show more compact nuclear morphology. This is
in line with the idea that the HEGs are the misaligned counterparts of
the face-on BLOs (B09+). On the other hand, LEGs show less extended
and compact morphologies (kpc or fraction of kpc) than HEGs. FR~Is are
generally more compact than FR~IIs, since all FRIs are LEGs. Generally,
H$\alpha$ and [O~III] show similar structures. However, there are
cases in which the morphologies of the two emissions are substantially
different. For example, the H$\alpha$ image of 3C~196.1 shows emission
extended $\sim$8 kpc to the northeast of the nuclear region which is
not observed in the corresponding [O~III] image. Interesting are the
cases of 3C~136.1 and 3C~196.1, which show two bi-conical structures
of emission in the central region. A careful spatial comparison
between the emission-line regions and the radio structures will be
addressed in a forthcoming paper.

\begin{figure}
\includegraphics[scale=0.43]{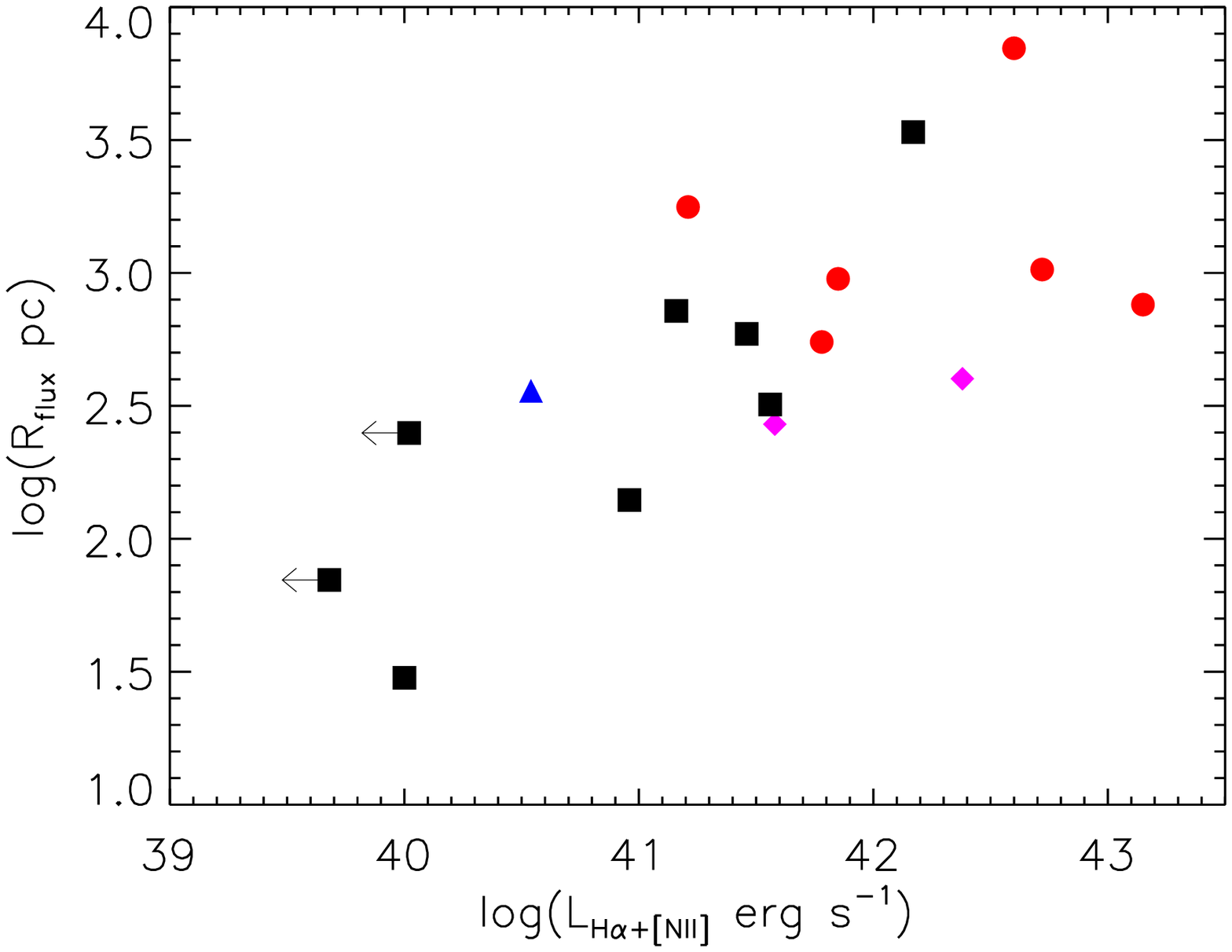}
\caption{Comparison between R$_{\rm flux}$ and H$\alpha$ luminosity
  for our sample. The black squares are LEG, the red circles
  are HEGs, the pink diamonds are BLOs and the blue triangle is the
  ELEG (3C~314.1).}
\label{size_lha}
\end{figure}

A straightforward way to measure the size of the NLR is to calculate
the so called `flux weighted radius' (R$_{\rm flux}$), using
H$\alpha$ images, defined as:

\begin{equation}  
R_{\rm flux}  = \frac{\rm \sum_{pix} (F_{pix} \times d_{pix})}{\sum_{\rm pix} {\rm F_{\rm pix}}}  
\end{equation}

where F$_{\rm pix}$ and d$_{\rm pix} $ are the flux of each pixel and
the distance of the pixel to the center of the galaxy, indicated with
a cross in Figure~\ref{panel}. The values, presented in
Table~\ref{detect2}, range from 0.03 kpc in the case of 3C~129 to 7.00
kpc in the case of 3C~180, with a mean value for the whole sample of
0.55 kpc. The typical uncertainty on R$_{\rm flux}$ is $\sim$22\%.
HEGs have larger values of R$_{\rm flux}$ than LEGs, on average by a
factor of $\sim$2.3, but still consistent with each other within
  the errors. Figure~\ref{size_lha} shows how the R$_{\rm flux}$
scales with the line luminosity for our sample. A clear trend R$_{\rm
  flux}$ -- L$_{\rm H\alpha}$ is present, indicating that higher
  luminosity targets have larger ELRs.

\begin{figure}
\includegraphics[scale=0.35,angle=90]{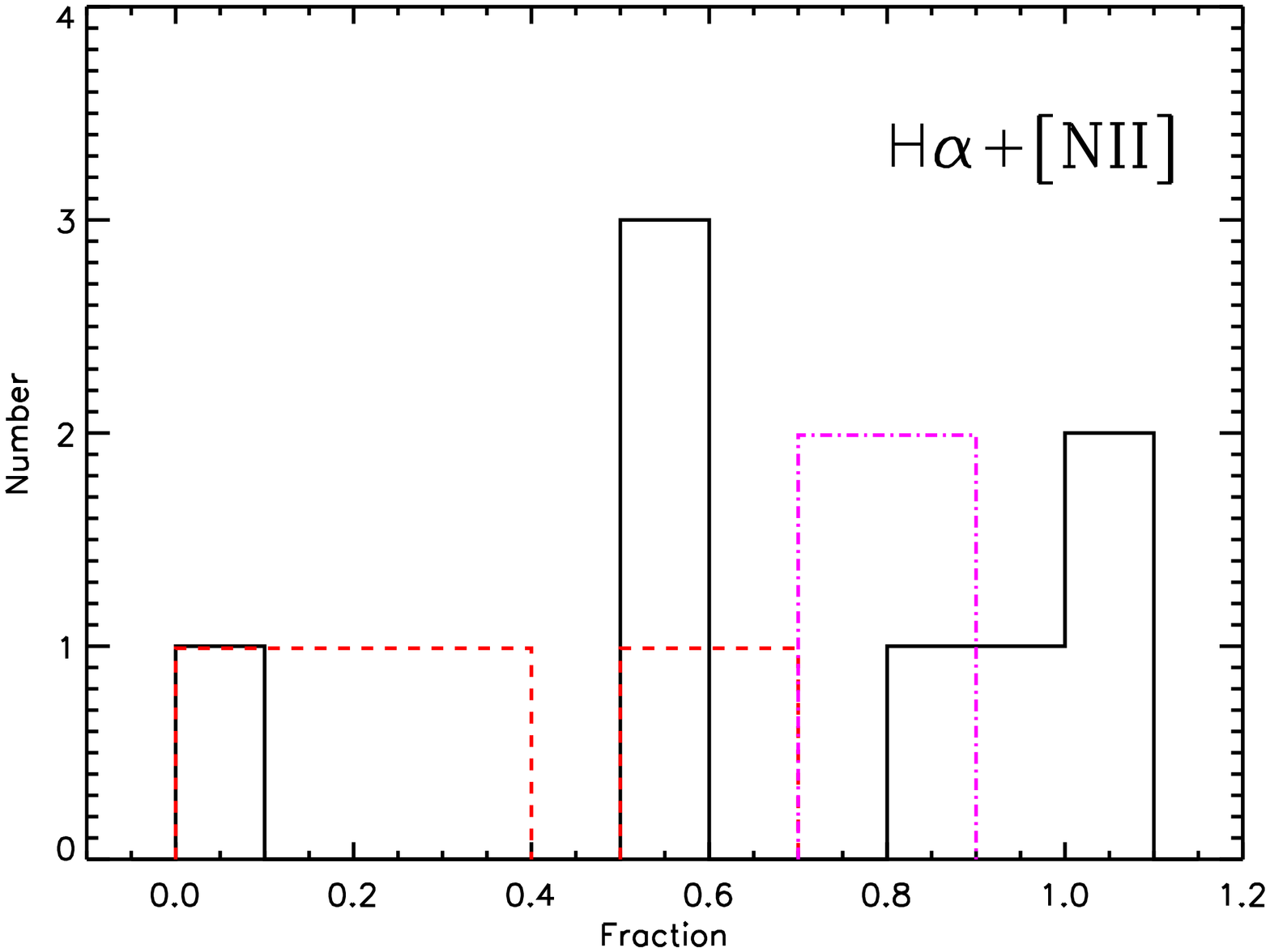}\\
\includegraphics[scale=0.35,angle=90]{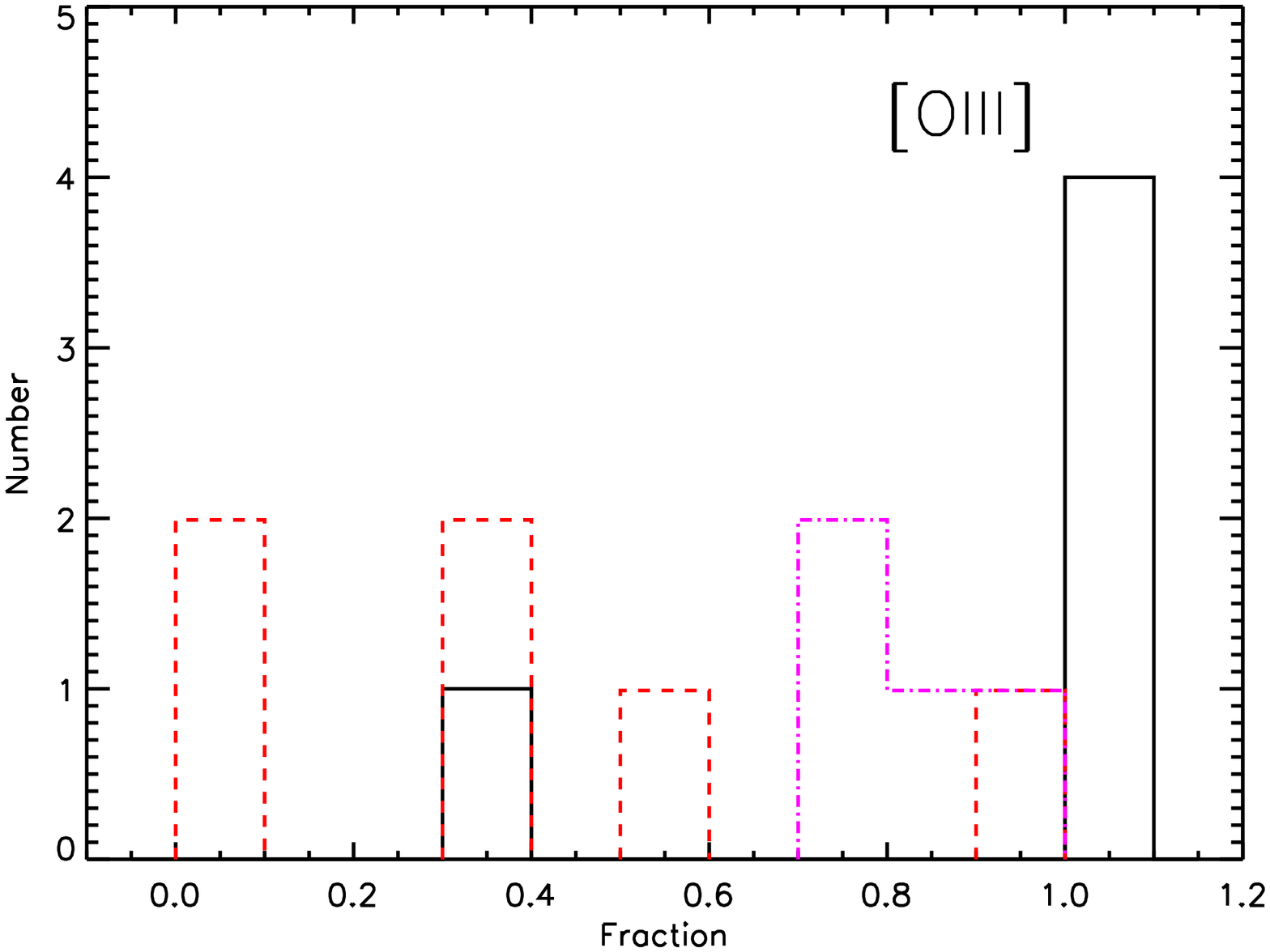}
\caption{Histogram showing the detected fraction of the total H$\alpha$
  luminosity (upper panel) and [O~III] luminosity (lower panel) within
  the central kpc. The black solid histogram stands
  for LEGs, the dashed red for HEGS, and the dot-dashed pink for
  BLOs.}
\label{fraction}
\end{figure}

To give an idea of the concentration of the emission,
Figure~\ref{fraction} depicts the fraction of the H$\alpha$ and the
[O~III] emission contained within the central kpc. The majority of the
sample exhibits the line emission concentrated mostly in the inner 1
kpc. The fraction of the H$\alpha$ line within the central kpc to the
total H$\alpha$ emission range from 0.07 to 1, with a median value of
0.64.  In the case of [O~III] the range goes from 0.04 to 1.0, with a
median value of 0.87. The objects, which show large ELR emitting over
1 kpc, are HEGs, with the exception of one LEG, 3C~196.1. Regarding
the HEGs, the H$\alpha$ and [O~III] emitted in the central kpc is
roughly half than what is seen in LEGs in the same area. This means
that LEGs typically show more compact emission-line region than HEGs.

\subsection{The [O~III]/H$\alpha$ line ratio images}

\begin{figure}
\includegraphics[scale=0.4]{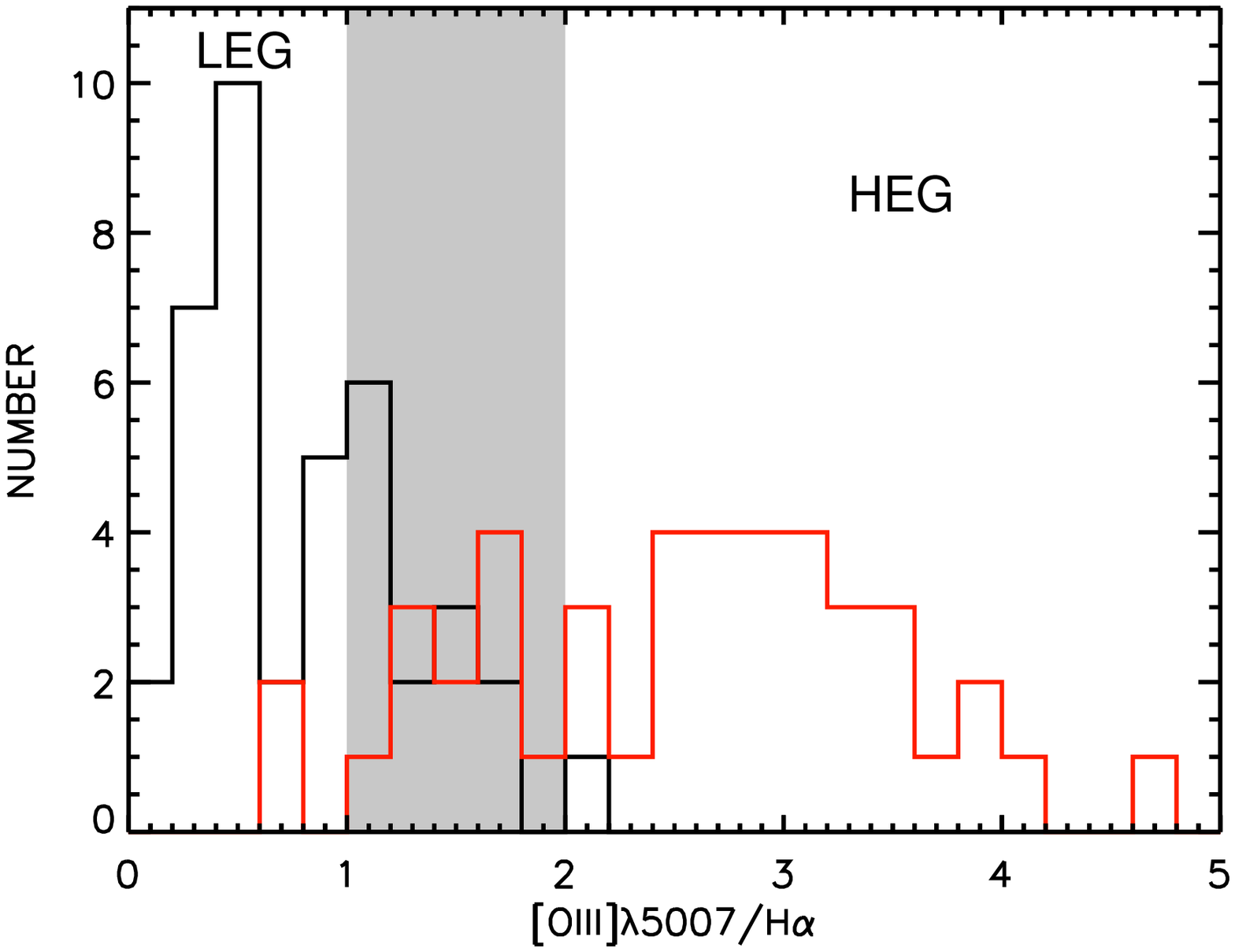}
\caption{Histogram showing the [O~III]/H$\alpha$ distribution for the
  113 3C radio sources in the \citep{buttiglione09} sample.  For the
  plot, BLOs were considered as HEGs. The black and red lines
  correspond to LEGs and HEGs respectively, and the grey-shaded region
  is the overlapping region where HEGs and LEGs share similar
  [O~III]/H$\alpha$ line ratios (1-2).}
\label{hist_LEG_HEG}
\end{figure}

\begin{figure*}
\includegraphics[scale=1.0]{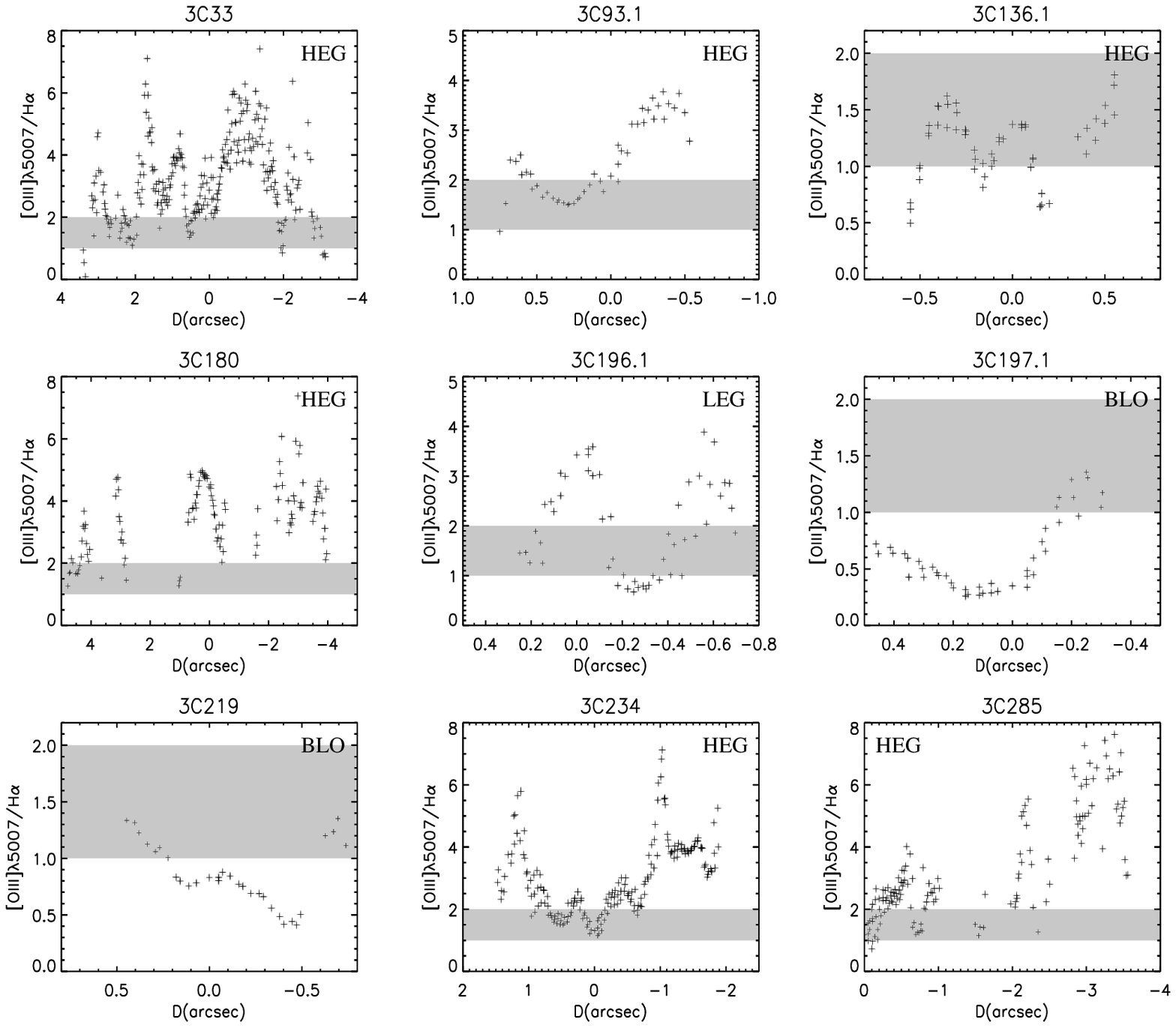}
\caption{The [O~III]/H$\alpha$ line ratios plotted against the
  distance from the nuclear region along the optical axis indicated in
  Figure \ref{panel}. We used the values in Column 6 and 7 in
  Table~\ref{detect2} to account for [O~III]$\lambda$4959 and
  [N~II]$\lambda\lambda$6949,6983 contribution to the flux in the
  band. The gray area corresponds to the range of
    [O~III]/H$\alpha$ ratios, where LEG and HEG optical
    classifications overlap (see Fig. \ref{hist_LEG_HEG}.}
\label{ratiovsdistance}
\end{figure*}

The third column in Figure~\ref{panel} depicts the log of the
[O~III]/H$\alpha$ emission line ratios for those galaxies with
extended emission and/or sufficiently high ($>$ 5) signal-to-noise
(3C~33, 3C~93.1, 3C~136.1, 3C~180, 3C~196.1, 3C~197.1, 3C~219, 3C~234
and 3C~285). These sources are all FR~IIs. They are all classified as
HEGs (including 2 BLOs) except for one LEG (3C~196.1).

In the emission-line diagnostic diagrams
\citep{kewley06,buttiglione09}, a net separation between LEGs and HEGs
occurs by combining several narrow optical lines (e.g.,
[O~III]/H$\beta$, [N~II]/H$\alpha$). Since we only studied [O~III] and
H$\alpha$, we cannot fully set up such diagnostics. However, these two
lines can still separate the two classes although less sharply than
the whole set of line ratios. Therefore, we used the [O~III]/H$\alpha$
ratios presented in the spectroscopic study of 113 3C radio sources
(B09+) in order to estimate the 'typical' line ratio values associated
with HEGs and LEGs.  Note that those ratios are taken from long-slit
nuclear spectra.  Figure~\ref{hist_LEG_HEG} shows the histogram of the
[O~III]/H$\alpha$ of the 3C sample, where BLOs are naturally
considered as HEGs. The black and red lines correspond to LEGs and
HEGs respectively. As described above, a division between these two
classes appears in the [O~III]/H$\alpha$ ratio distribution, but with
a broad overlapping region, shown by the grey-shaded area.  HEGs and
LEGs share similar [O~III]/H$\alpha$ line ratios in the range of
values 1 $-$ 2 (0$-$0.3 in logarithmic scale). We use this range of
ratios as reference to separate low- and high-excitation line ratios
for our maps.

The emission line ratios maps in Figure~\ref{panel} (third column)
reveal that the [O~III]/H$\alpha$ ratio may change across the ELR
(between 0.2 and 5) from a LEG to HEG classification and
viceversa. The ratio values shown to the right of the images are not
corrected for either the [N~II]$\lambda\lambda$6549,6583 or the
[O~III]$\lambda$4959 contribution.

To quantitatively study the [O~III]/H$\alpha$ ratio variation across
the galaxy, we opted for analyzing the line ratios along a
preferential direction within the galaxy, for simplicity. This
direction is the optical axis (marked as line and labelled as OA in
Figure~\ref{panel}, central panels), defined as the axis crossing the
center of the galaxy, along which the optical emission is more
extended in either the [O~III] or the H$\alpha$ images.
Figure~\ref{ratiovsdistance} plots the [O~III]/H$\alpha$ line ratio
values against the distance from the center of the galaxy, indicated
with a cross in Figure~\ref{panel} along the optical axis.  In
Section~A.1 we performed a test to confirm the reliability of the
plots showed in Figure~\ref{panel} and Figure~\ref{ratiovsdistance}.
The plots show a intense scatter pattern of the intensity ratio across
the galaxy length, which reveal the large variation of the physical
condition present in the ELRs on short scales.

When generating the plots in Figure~\ref{ratiovsdistance}, we correct
the emission line ratio for [O~III]$\lambda$4959 and
[N~II]$\lambda\lambda$6949,6983 contribution to the flux in the band
(see correction factor in Table~\ref{detect2}), but not for
reddening. In order to account for the effects of reddening, we
examine the optical HST (F555W and F702W) WFPC2 images
\citep{martel99,dekoff96} to search for dust features across the
galaxies. We find that such components are evident in three of the
galaxies in our sample, 3C~33, 3C~136.1 and 3C~285.

To further investigate the effects of dust on our emission line maps,
we used the H$\alpha$ and H$\beta$ flux values from B09+ to calculate
the H$\alpha$/H$\beta$ ratio, which is sensitive to
reddening. Assuming Case B recombination theory, this line ratio is
$\gtrsim$3 if reddening is present. We found line ratios significantly
higher than this value for three of the galaxies in our sample
(3C~136.1, 3C~285 and 3C~319) due to the dimmed H$\beta$. For 3C~319
the spectrum is intrinsically weak and H$\beta$ and [O~III] are not
detected, probably due not to absorption.  Since dust attenuation is
more important at shorter wavelengths, the [O~III]/H$\alpha$ line
ratios presented in that figure for 3C~33, 3C~136.1, and 3C~285 can be
considered as lower limits.

Two of the nine RGs included in Figure \ref{ratiovsdistance} (3C~197.1
and 3C~219) are classified as BLOs.  Taking into account the
contribution of the broad line in our emission-line images, estimated
from nuclear spectra, their line ratios at the central location are
underestimated by a factor of $\sim$1.7 - 4.

The most interesting result is the variety of morphologies and sharp
variations in the [O~III]/H$\alpha$ line ratio images, which
are mainly observed in HEGs. Interesting is the case of 3C~33, two
regions with [O~III]/H$\alpha$ line ratios of 4-8 are observed
$\sim$1.5 $\arcsec$ ($\sim$1.7 kpc) to the north-east and south-west
of the nuclear region. 3C~180 shows line ratios of up to 8 spread
through the filaments of emission observed 2-5$\arcsec$ (7-18 kpc) to
the north and south of the nucleus. Another interesting case is
3C~234, where the highest line ratio values (6-8) are located at
1-2$\arcsec$ (3-6 kpc) to the west from the nuclear region.

Overall, HEGs show the highest line ratios ([O~III]/H$\alpha$ $>$2)
and largest scatter even at large distance from the galaxy center
(Figure~\ref{ratiovsdistance}). For the two BLOs and one LEG, the
ratio variations are shallower.

\subsection{Luminosity correlations}

\begin{figure}
\includegraphics[scale=0.4]{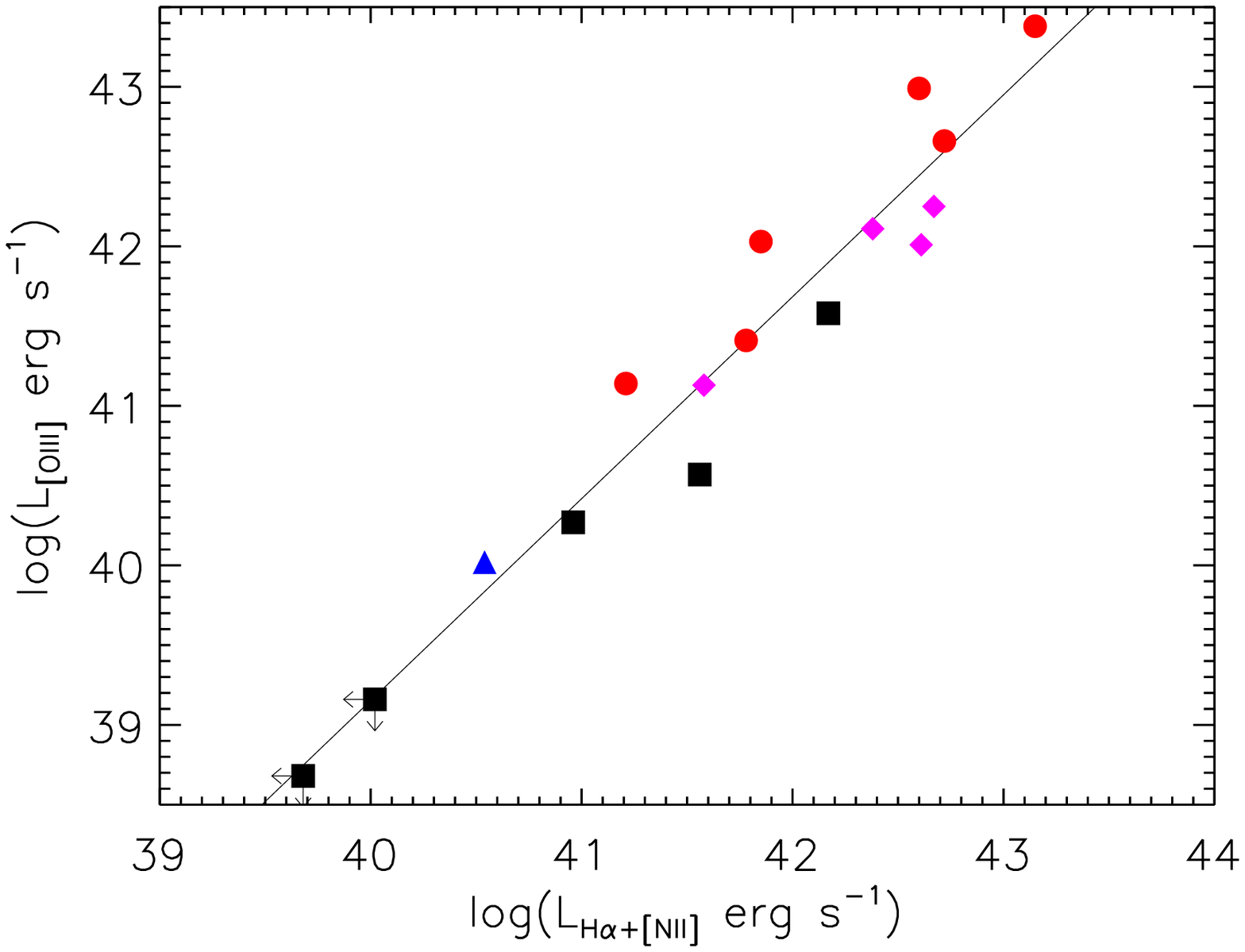}
\caption{Log of the [O~III] luminosity plotted against the log of the
  H$\alpha$ luminosity for the 16 galaxies in our sample both
  measurements are available. The symbols are coded as in
  Fig.~\ref{size_lha}.}
\label{HavsOIII}
\end{figure}

In Figure~\ref{HavsOIII} we compare the H$\alpha$ and [O~III]
luminosities for the galaxies in our sample. A tight correlation is
observed between the two emission lines in the form L$_{\rm
  [O~III]}$$\propto$ (L$_{\rm H\alpha + \rm
  [N~II]}$)$^{1.26\pm0.10}$. We checked the statistical significance
of such a trend by using a censored statistical analysis (ASURV;
\citealt{lavalley92}) which takes into account the presence of upper
limits. The rms of the residuals and the correlation coefficient are
0.35 dex and 0.96 respectively.  If we correct the broadband images
considering the [N~II] and the second [O~III] contribution from B09+
(Tab~\ref{detect2}), a correlation is still present, with slope of
1.19$\pm$0.14. The rms of the residuals and the correlation
coefficient are 0.50 dex and 0.95 respectively.  LEGs and HEGs lie
  on the same relation.

\begin{figure}
\begin{tabular}{c}
\includegraphics[scale=0.4]{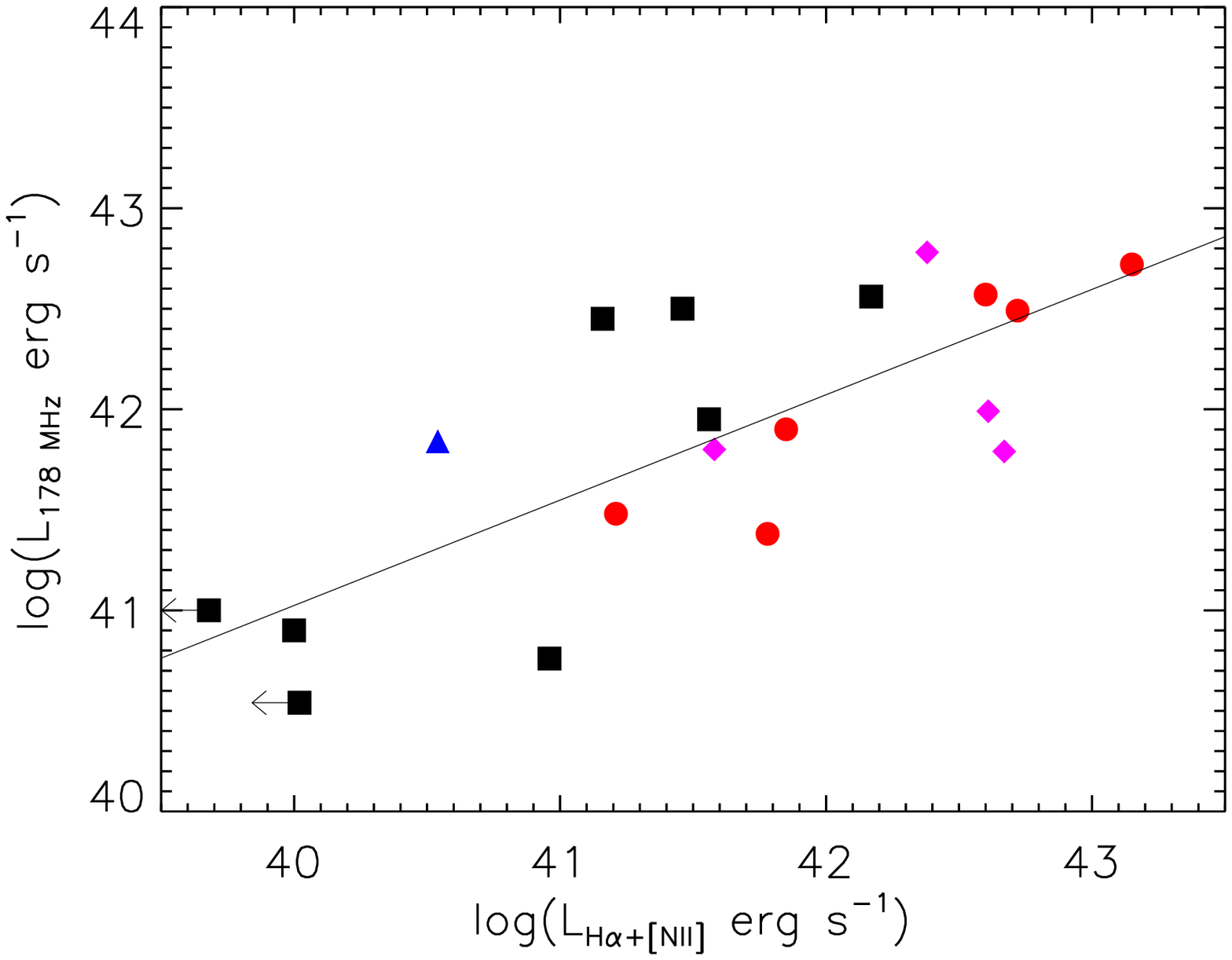}\\
\includegraphics[scale=0.4]{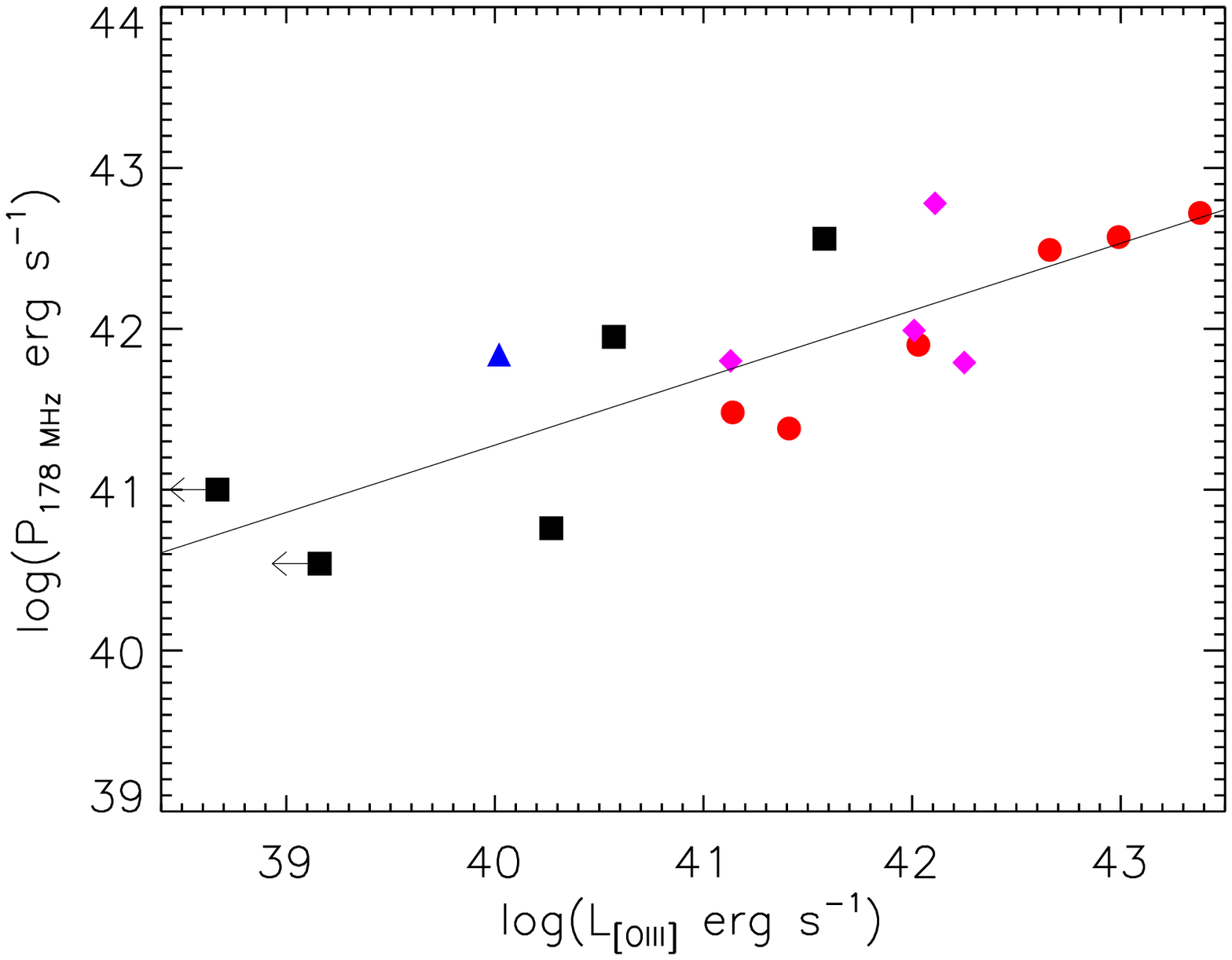}
\end{tabular}
\caption{Radio power at 178-MHz plotted against the [O~III]
  (upper panel) and the H$\alpha$ (lower panel)
  emission line luminosities. Symbols are the same as in Figure
  \ref{size_lha}.}
\label{optvsradio}
\end{figure}

We also used our {\it HST}-ACS observations to test the already known
correlation between the optical emission line luminosities and the 178
MHz radio power
\citep{Baum89a,Baum89b,Rawlings89,Rawlings91,Morganti92,Zirbel95}.
This is interpreted as a link between the AGN continuum emission and
the mechanism of radio jet production \citep[see][for a detailed
  discussion]{Tadhunter98}. Figure~\ref{optvsradio} shows the [O~III]
and the H$\alpha$ luminosities plotted against the 178 MHz radio power
(Table~\ref{multiband}). We found correlations in the form L$_{\rm
  [O~III]}$$\propto$ (L$_{\rm 178 MHz}$)$^{0.52\pm0.3}$ and L$_{\rm
  H\alpha} \propto$ (L$_{\rm 178 MHz}$)$^{0.70\pm0.2}$, which are
consistent with the correlations in the literature. The rms of the
residual and the correlation coefficient are 0.70 dex and 0.81 for
[O~III] and 0.60 dex and 0.76 in the case of the H$\alpha$
emission. When we correct for [N~II]$\lambda\lambda$6549,6583 and
[O~III]$\lambda$4959 contribution to the flux within the band, the
correlation is still present with slopes of 0.53$\pm$0.2 and
0.70$\pm$0.2 for the [O~III] and the H$\alpha$ emission lines
respectively. While HEGs and LEGs follow a common relation in
Figure~\ref{optvsradio}, \citet{buttiglione10} found different
relations for the two classes. Although this discrepancy could be
  caused by the different integration area between our HST data and
  their nuclear slit spectra, the results of \citet{buttiglione10} on
  the almost-complete 3CR sample are, however, statistically stronger
  than what we found for 19 sources in this work.

\begin{figure}
\begin{tabular}{c}
\includegraphics[scale=0.4]{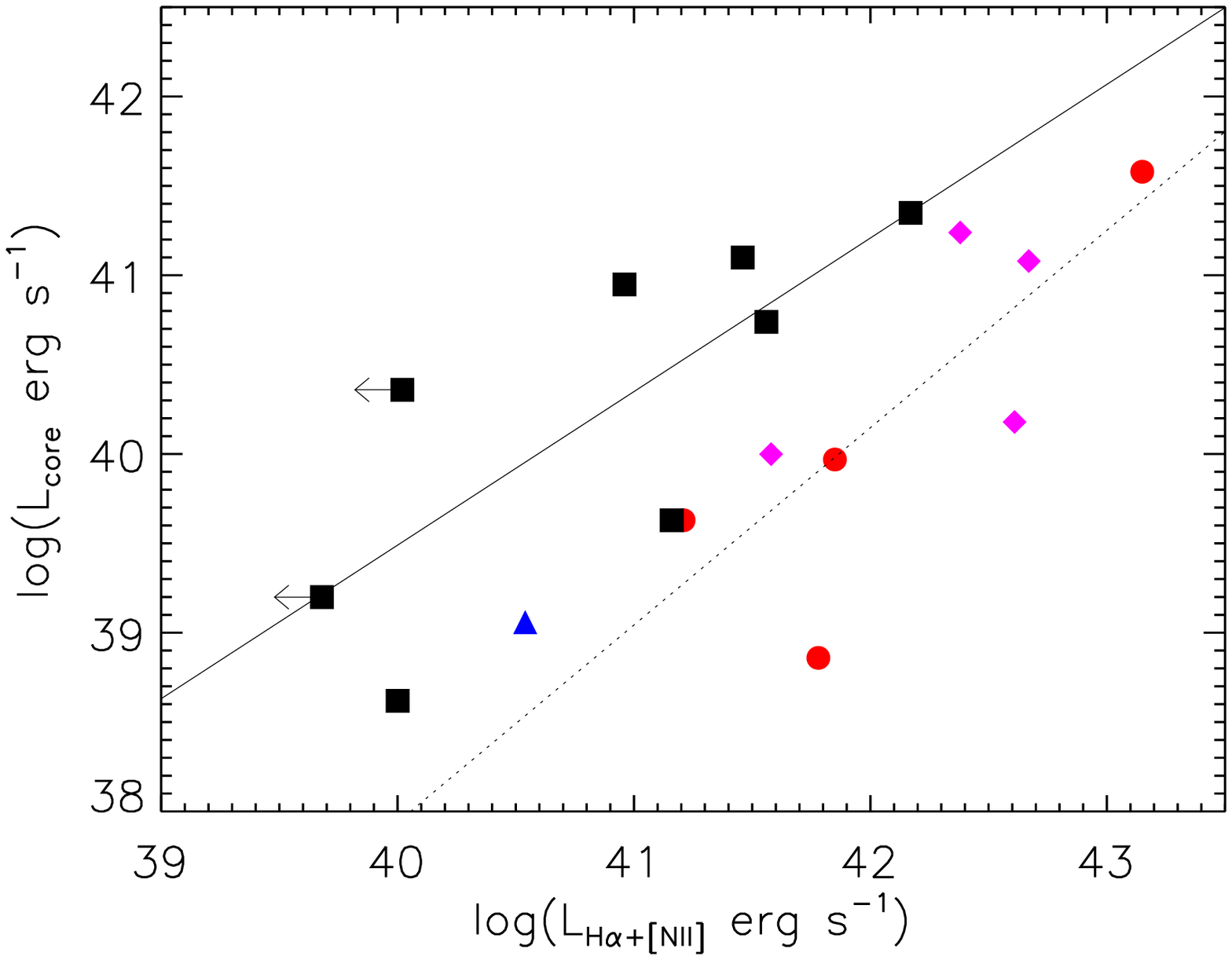}\\
\includegraphics[scale=0.4]{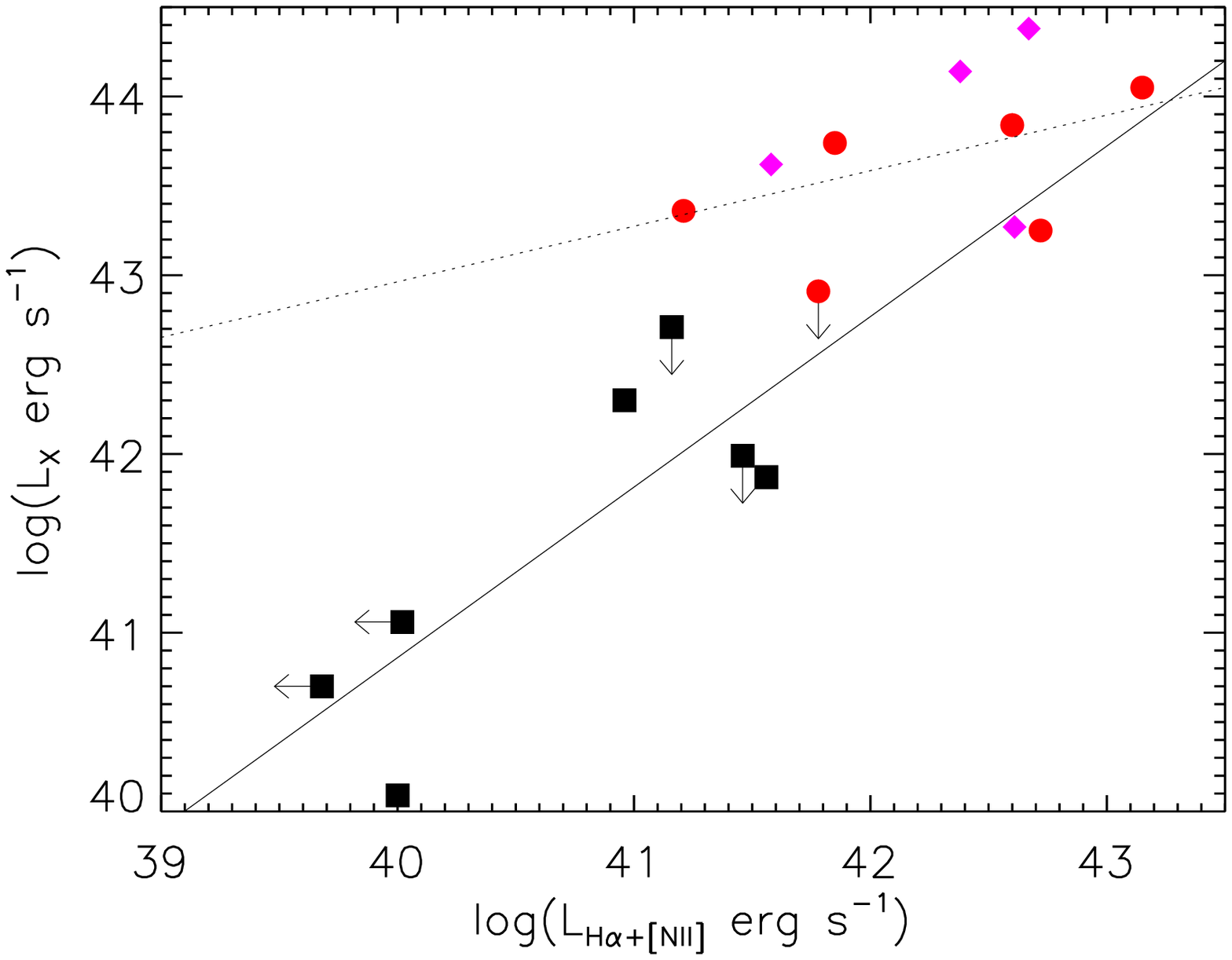}
\end{tabular}
\caption{Radio core power at 5 GHz (upper panel) and X-ray 
    intrinsic luminosities (2-10 keV) (lower panel) plotted 
    against H$\alpha$ emission line luminosities. Symbols are the
  same as in Figure \ref{size_lha}. The solid and dotted lines
    represent the best fits for LEGs and HEGs, respectively.}
\label{optvsnuclei}
\end{figure}

Figure~\ref{optvsnuclei} provides an interesting result. It shows the
5-GHz radio core power (upper panel) and 2-10 keV X-ray nuclear
luminosities (lower panel) in relation to the H$\alpha$
luminosities of our sample. The two classes of objects, LEGs and HEGs,
follow different linear correlations in these diagrams, in agreement
with the results from \citet{buttiglione10}. In fact, the probability
that the two classes lie on the same correlations in the
radio-H$\alpha$ and X-ray-H$\alpha$ planes is 10\% and 32\%,
respectively. In the radio-core panel, the linear relations for LEG
and HEG appear parallel (L$_{core} \propto$ L$_{\rm
  H\alpha+[N~II]}$$^{\alpha}$, where $\alpha$ is 0.86$\pm$0.15 for LEG
and 1.1$\pm$0.2 for HEG and similar rms of $\sim$0.7 dex). The
probabilities that the LEG and HEG radio--line correlations are
fortuitous are, respectively 0.0006 and 0.0003. In the X-ray-line
panel, the relations for the two classes are still distinct (L$_{X}
\propto$ L$_{\rm H\alpha+[N~II]}$$^{\alpha}$, where $\alpha$ is
0.3$\pm$0.2 for LEG and 0.95$\pm$0.30 for HEG and similar scatter of
$\sim$0.7 dex). The probabilities that the LEG and HEG X-ray--line
correlations are fortuitous are, respectively 0.0008 and 0.0005. By
correcting the line luminosities for [N~II]$\lambda\lambda$6549,6583
and [O~III]$\lambda$4959, the luminosity correlations are still valid
with scatters and slopes consistent with the values reported above.

To test the correctness of the radio and X-ray luminosity correlations
we found, we searched for the equivalent correlations in the flux-flux
parameter space. The flux-flux correlations are weaker than the
luminosity correlations, but still statistically valid
($<$9$\times$10$^{-4}$ of probability that a fortuitous correlation is
present), given that the flux data scatters in the plots are slightly
larger than those observed in Figures \ref{optvsradio} and
\ref{optvsnuclei}. In conclusion, all the correlations support the
idea that the total amount of line emission produced on large scales
for the two optical classes, LEGs and HEGs, depend on their different
physical properties at nuclear scale.

\subsection{Photon budget, mass and covering factor of the ionized
  gas}

\begin{table}
\begin{center}
  \caption{Properties of the ionizing source and ionized gas}
\begin{tabular}{lccccc}
  \tableline\tableline
  Name & Q$_{\rm ion}$    & $\Omega_{ELR}$ & n$_{e}$ & M$_{ion}$ \\ 
  3C    & photon s${-1}$&                & cm$^{-3}$& M$_{\odot}$\\ 
  (1)&(2)&(3)&(4)&(5) \\
  \tableline
  33.0 & 8.0$\times10^{51}$  & 0.39  & 267 & 6.49  \\
  40   & 1.7$\times10^{49}$  & $<$0.62 & --& --\\
  78   & 3.2$\times10^{50}$  & 0.28   & --& --\\
  93.1 & 2.6$\times10^{51}$  & 1.12  & 658 & 7.08 \\
  129  & 2.0$\times10^{48}$  & 1.03   & 606 & 4.58 \\
  132  & $<$3.7$\times10^{51}$ & $>$0.08   & 847& 5.89 \\
  136.1&$<$1.2$\times10^{51}$ & $>$0.51   & 606& 5.92 \\
  180  &1.0$\times10^{52}$ & 0.40   & -- & -- \\
  196.1& $-$              &   --        & 471 & 5.54 \\
  197.1&5.6$\times10^{51}$ &  0.37   & 806& 5.28 \\
  219  &2.0$\times10^{52}$ & 0.32  & 606& 6.23  \\
  227.0&2.6$\times10^{51}$  & 1.8  & 606& 5.79 \\
  234  &1.6$\times10^{52}$  & 0.9  & 328 & 7.70 \\
  270  & 9.4$\times10^{49}$ & $<$0.05       &--   &  --\\
  285  & 3.6$\times10^{51}$  & 0.15  & 10& 7.32\\
  314.1& $-$    & --   & -- & -- \\
  319  & $<$7.5$\times10^{50}$& $>$0.19 & 451 & 5.78  \\ 
  388  &1.2$\times10^{50}$ &  1.05 & 381 & 5.74  \\
  390.3& 3.6$\times10^{52}$& 0.33 & 606 & 6.11 \\
  \tableline
\end{tabular}
\label{Covering}
\tablecomments{Col (1): 3C Name. Col (2): number of ionizing photons
  Q$_{ion}$.  Col (3): covering factor.  Col (4): electron
  densities estimated using the [S~II]6716/6731 line ratios from
  integrated optical spectra \citep{buttiglione09,buttiglione10}.  Col
  (5): ionized gas mass.}
\end{center}
\end{table}

In this section we calculate the ionizing photon budget for the
emission-line regions and estimate the ionized gas masses and their
covering factors for the RGs in our sample.

Assuming isotropic radiation from the central source, the number of
ionizing photons, Q$_{\rm ion}$, is obtained by integrating the luminosity
density over the whole spectrum above ${\rm \nu_{o}}$, which is the
frequency corresponding to the ionization potential of hydrogen i.e:

\begin{equation} 
{\rm Q_{ion}} = \int_{\nu_0}^{\infty} \frac{\rm L_{\nu}}{h\nu} d\nu 
\end{equation}

If we now assume a power law spectral shape in the form L$_{\nu}$ = C~(${\rm
  \nu}$/${\rm \nu_{ref}}$)$^{-\alpha}$, where ${\rm \nu_{ref}}$ is any frequency
for which the corresponding luminosity is known, equation 3 can be written as:

\begin{equation} 
{\rm Q_{ion}} = \frac{\rm C}{\rm h\alpha}(\frac{\rm \nu_0}{\rm \nu_{ref}})^{-\alpha} 
\end{equation}

where the constant $C$ is the normalization of the spectrum. The
ionizing photons are produced in two main regions of the
electromagnetic spectrum, in the UV-optical and X-ray range. The
UV-optical nuclear luminosities of HEGs
\citep{chiaberge02,chiaberge02b} would significantly underestimate
their real photon budget because their nuclei are obscured, partially
obscured or dominated by scatter emission
\citep{baldi10b,baldi13}. Conversely, hard-X-ray luminosity (2-10 keV)
is a better proxy to estimate the number ionizing photons than the
UV-optical band, because less effected by obscuration.

Therefore, we derived the Q$_{\rm ion}$ from the X-ray wavelengths. In
order to determine the value of the constant $C$, and hence the number
of ionizing photons, we use the 2-10 keV X-ray luminosities and photon
indexes $\Gamma$ ($\Gamma$ = $\alpha$ + 1) for the galaxies in our
sample from Torresi et al. (in prep) (using {\it Chandra} data,
  Table~\ref{multiband}) as follows:

\begin{equation} 
{\rm L_{X-ray}} = \int_{2keV}^{10keV} {\rm L_{\nu}}d\nu 
\end{equation}

\begin{figure}
\begin{tabular}{c}
\includegraphics[scale=0.4]{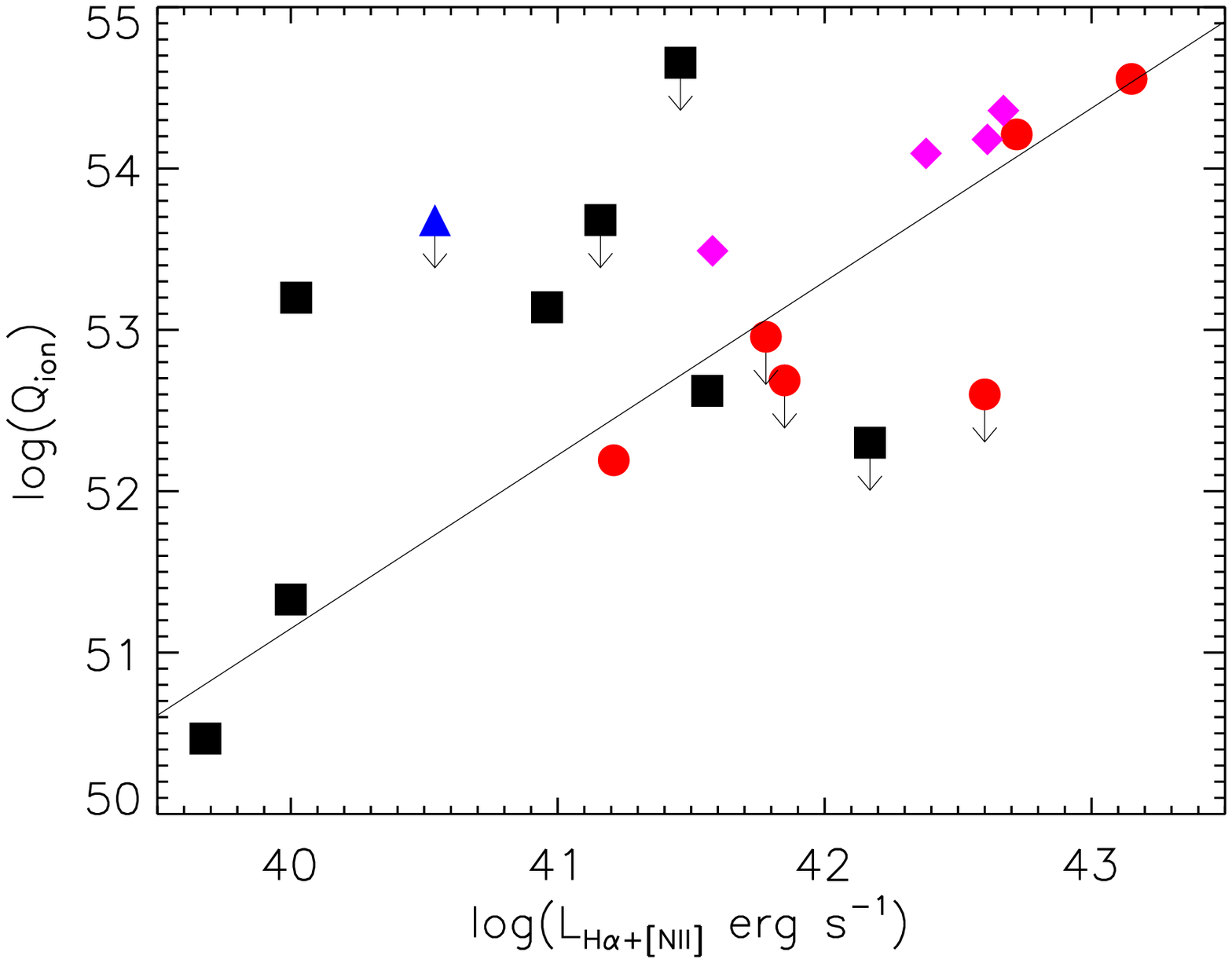}
\end{tabular}
\caption{H$\alpha$ luminosity vs number of ionizing photons
    estimated in the X-ray (2-10 keV) band. Symbols are the same as in
    Figure \ref{size_lha}. Using a censored statistical analysis, a
    tentative trend is observed whose slope coefficient is
    1.08$\pm$0.25.}
\label{LHaQion}
\end{figure}

The number of ionizing photons in the X-ray band are shown in
Table~\ref{Covering}. Figure~\ref{LHaQion} displays the H$\alpha$
luminosities versus the total number of ionizing photons, Q$_{ion}$. A
monotonic, weak trend between the two quantities appears (probability
that a fortuitous correlation appears is 0.0035). This indicates that
the AGN primarily provides the ionizing photons to account for the
emission-line luminosities for our sample of RGs.

We estimate the covering factors of the ELRs, $\Omega_{\rm ELR}$, by
using the L$_{H\alpha}$ from our maps. In the case of low-density NLR
clouds, in case-B recombination, the number of ionizing photons can be
related to the H$\alpha$ luminosity in the form \citep{osterbrock89}:

\begin{equation}  
 \frac{\rm L_{H\alpha}}{\rm 2 h\nu_{\alpha}} =   \frac{\rm \Omega_{\rm ELR} Q_{ion}}{\rm 2.2}
\end{equation}

Assuming a bolometric 2-10 keV correction factor of 30
\citep{vasudeven10,lusso12} to correct the ionizing photon budget, we
measure the covering factors of our sample, presented in
Table~\ref{Covering}.  ${\rm \Omega_{\rm NLR}}$ range between
$\sim$0.05 and 1.8 with a mean value of 0.66. Four sources (3C~93.1,
3c~129, 3C~227, 3C~388) show covering factors larger than unity. LEGs
show slightly smaller mean covering factors (0.60) than HEGs and BLOs
(0.71), but the difference is not statistically significant (P $<$
90\%).

 As discussed in \cite{osterbrock89}, we can
use the H$\alpha$-line luminosity to estimate the mass of the ionized
gas for the galaxies in our sample using the expression:
 
\begin{equation}  
 M_{ion} = \frac{\rm L_{H\alpha}m_p}{\rm \alpha^{eff}_{H\alpha}h\nu_{\alpha}n_e}  
\end{equation}

where m$_{\rm p}$ is the the mass of the proton, ${\rm
  \alpha^{eff}_{H\alpha}}$ is the H$\alpha$ recombination coefficient
(1.17 $\times$ 10$^{-13}$ cm$^{3}$ s$^{-1}$) and n$_{e}$ is the
electron density. To estimate n$_{e}$, we use the
[S~II]$\lambda\lambda$6716,6731 emission line ratio\footnote{We used
  the {\it TEMDEN} IRAF routine to compute electron density from
  diagnostic line ratios.}, obtained from the optical spectra
(B09+). Table~\ref{Covering} presents n$_{e}$ for the galaxies in our
sample and the corresponding ionized gas mass estimates, M$_{ion}$. In
the cases of 3C~78 and 3C~270 it is not possible to obtain adequate
[S~II]$\lambda\lambda$6716,6731 measurements. For 3C~40, 3C~180 and
3C~314.1 the [S~II]6716/6731 line ratios are outside the low density
limit (1.52). We obtain a wide range of electron densities (10 -- 806
cm$^{-3}$) with a median value of 606 cm$^{-3}$.  These values give
ionized gas masses ranging from 3.8 $\times$ 10$^{4}$ to 5.0 $\times$
10$^{7}$ M$_{\odot}$, with a mean value of 6.6$\times$10$^{6}$
M$_{\odot}$. The interesting result is that HEGs have more
  massive ELRs than LEGs by a factor of $\sim$8.5. Note that the [S~II]
ratio is estimated in the nuclear region and probably not
representative of the conditions in the outer region of the ELR.
Consequently, for a more tenuous ELR, this approach might lead to an
underestimation of the actual ionized gas masses.

\subsection{Radio -- optical alignment}

In this section we qualitatively study the radio and optical alignment
for our sample. A detailed spatial superimposition of the radio and
emission-line region maps will be presented in a forthcoming paper.

The black-solid lines drawn in the H$\alpha$ images in Figure
\ref{panel} show the orientation of the optical axis. It is defined as
the angle between the optical axis (defined in Sect 4.4) and the North
direction. There are eight sources (3C~129, 3C~132, 3C~219, 3C~270,
3C~314.1, 3C~319, 3C~388 and 3C~390.3) in which the relatively low
signal-to-noise and/or the amorphous morphology of the detected
emission makes the definition of an optical axis rather
arbitrary. Therefore, for these eight sources no optical axis is shown
in Figure~\ref{panel}. In addition, the radio axes of the galaxies
based on high-resolution VLA images (jet position angle,
\citealt{dekoff96,martel99,privon08} and references therein) are
plotted in Figure \ref{panel} onto the [O~III] images.

\begin{figure}
\includegraphics[scale=0.4]{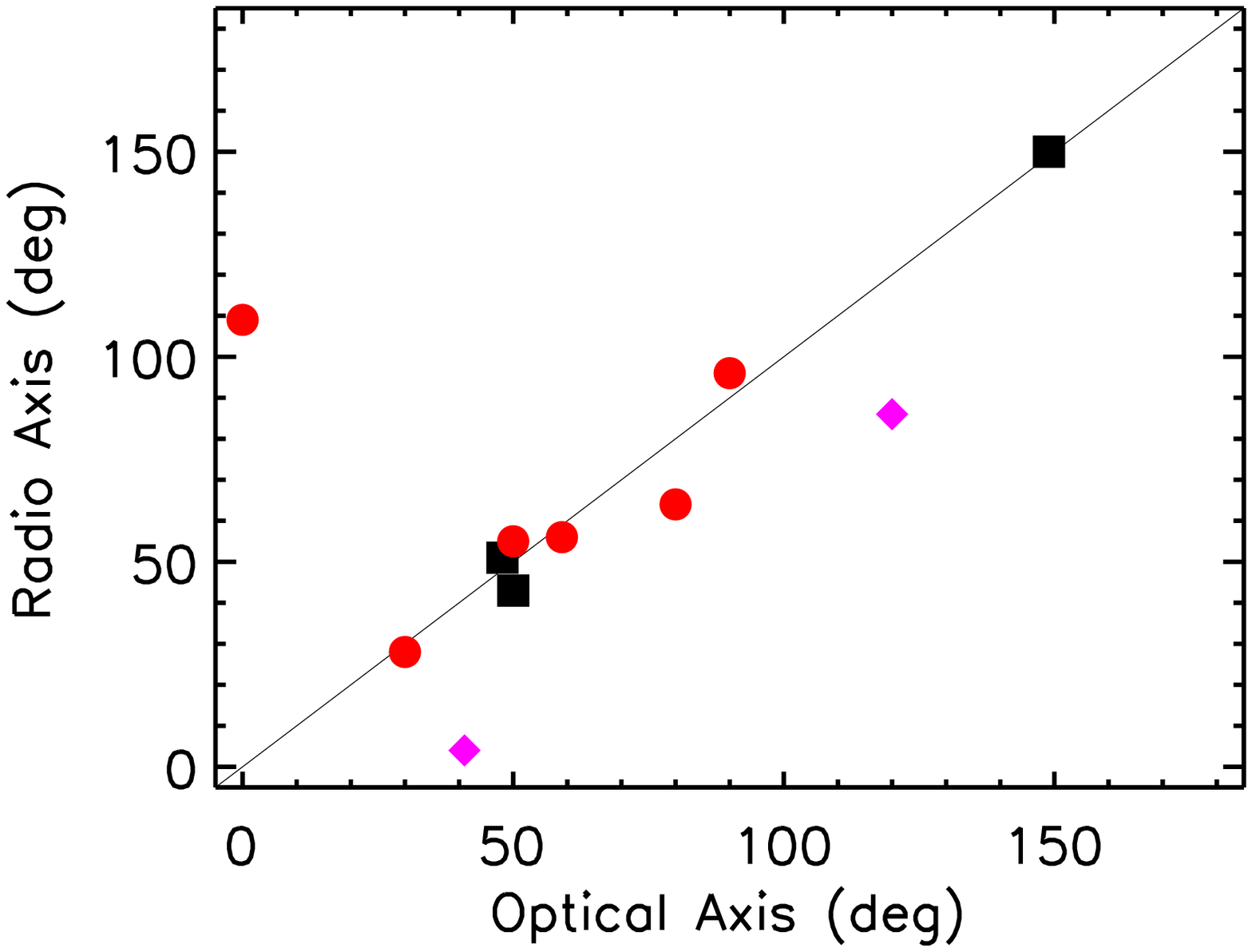}
\caption{Radio axis plotted against the optical axis (N is 0$^{\circ}$
  and E is 90$^{\circ}$, see text for details). The values are in
  degree. The symbols are coded as in Fig.\ref{size_lha}.}
\label{radio-vs-optical}
\end{figure}

Figure \ref{radio-vs-optical} describes the radio axis plotted against
the optical axis for the eleven sources in our sample with the two
measurements available. The main result is that the two axes have
similar orientations. This suggests that the optical emission is
tightly related to the radio emission, as expected by the the
'alignment effect' observed in RGs. The source which shows the largest
radio/optical axis discrepancy is 3C~136.1 which shows a bipolar
emission-line structure, traversal to the jet.

\section{Discussion}

Our results on the HST images of local RGs point towards the active
nucleus as the main driver of the emission-line properties we
observe. We note a different behavior between the ELR properties of
the AGN populations, HEG and LEG, in our sample: larger sizes, masses,
luminosities in favor of HEGs  (Table~\ref{comparison} summarizes
  the mean values of the two classes for comparison). Generally for
  our radio sample, higher line luminosities are associated with
  larger ELRs, in agreement with the size-luminosity dependence found
  for Seyferts and QSOs \citep{fisher18}.

Morphologies and luminosities of the ELRs are consequences of two
factors: the source of ionizing photons and the available content of
cold gas in the ISM. These two elements are not completely disjointed
since AGN feedback on the host and gas accretion on the BH are
mutually regulated processes. In the next sub-sections we discuss the
topic of emission-line gas in RG, addressing the ionization mechanism
and the gas properties, in the framework of the two classes of
radio-loud AGN, LEG and HEG, which are known to correspond to two
different accretion modes, hot and cold, respectively
(e.g. \citealt{best12}).

\begin{table}
\begin{center}
\caption{Comparison between LEGs and HEGs}
\begin{tabular}{ccccc}
\tableline\tableline
quantity       & sample & LEG   & HEG  & units \\
\tableline
log L$_{\rm H\alpha}$ &  41.58$\pm$0.23  & 40.85$\pm$0.30  &  42.26$\pm$0.18 & erg s$^{-1}$  \\
log L$_{\rm [O~III]}$ &  41.31$\pm$0.34  & 39.96$\pm$0.51  &  42.11$\pm$0.23 &  erg s$^{-1}$  \\
R$_{flux}$          &   1.09$\pm$0.41  & 0.69$\pm$0.37 &   1.59$\pm$0.74  & kpc  \\
$\Omega_{\rm ELR}$   & 0.66$\pm$0.13  & 0.60$\pm$0.17  &   0.71$\pm$0.17  &  $-$ \\
log M$_{ion}$       &  6.82$\pm$0.21 & 5.67$\pm$0.21 &   7.00$\pm$0.25 & M$_\sun$\\
\tableline
\label{comparison}
\end{tabular}
\tablecomments{Mean values of the physical quantities with
    1$\sigma$ errors obtained for the entire sample and the two
    optical classes, LEGs and HEGs. We used the Kaplan-Meier estimator
    to derive the mean values by considering the censored data with
    ASURV package.}
\end{center}
\end{table}

\subsection{Ionization mechanism}

Early studies clearly show that the optical emission-line gas is
ionized by the nuclear continuum, whose properties are set by the
central engine (e.g. \citealt{Baum89b,Rawlings91}). The relation of
the line emission with the radio jet power indicates that its kinetic
energy deposited in the ELR has an important contribution in the
energy budget of the photoionization of the clouds, independently of
the AGN class. Linear correlations found between the emission-line
luminosity and the nuclear activity indicators, such as X-ray and
radio core luminosities (Figures ~\ref{optvsradio} and
\ref{optvsnuclei}), confirm this scenario. However, those correlations
clearly indicate a discrepancy between LEGs and HEGs. The HEG
  population shows an excess of emission line luminosity of a factor
  $\sim$10 with respect to the LEGs, when matched in radio core
  luminosities. Such a line excess is indicative of an additional
  ionizing component to the radio jet.  Instead, in the X-ray--line
  plane, the two classes occupy different regions of the diagram, with
  HEGs more luminous in general than LEGs, suggesting two different
  main ionizing mechanisms responsible to the line properties of the
  two populations.

Many recent multi-band studies indicate a bi-modality in the accretion
modes for RL AGN. In turn, such a behavior corresponds to a
bi-modality in the '{\it ionization modes}'. For the LEG population
(the hot-gas accretors), the synchrotron radio emission from the jet
is enough to account for their multi-wavelength properties: X-ray, UV,
optical, IR and line luminosities correlate with the radio core, which
is a proxy of the jet energetics (e.g.,
\citealt{balmaverde06a,balmaverde06core,chiaberge99,chiaberge02,capetti05b,baldi10b,Hardcastle09}
and other references therein). Conversely, HEGs (the cold-gas
accretors) systematically exhibit an excess with respect to the
synchrotron-jet component, which is ascribed to the standard accretion
disk to account for their higher multi-band and line luminosities than
the LEGs. Our results point to a linked bi-modality, accretion
vs. ionization. On one hand, the picture of a radio jet illuminating
the NLR of LEGs fits the previous multi-band studies. On the other
hand, the accretion disk provides further ionizing photons to
irradiate the ELR of HEGs.

Once the radiation field is produced either by the jet or the
accretion disk, the ISM in the host galaxies reprocesses the ionizing
light into emission lines depending on their available cold gas
amount. The dust and cold gas present in a large amount in HEGs are
visible in the optical band and detected either in the IR band or in
molecular lines (10$^7$-10$^8$ M$_{\odot}$,
\citealt{baldi08,smolcic11,tadhunter14,westhues16}). These are usually
tracers of the location of the emission-line gas
\citep{Dicken09}. Conversely, LEGs are typically hosted in red
gas-poor, passively evolving ETGs
\citep{baldi08,smolcic09,prescott18}, which justify their small sizes
and moderate masses of the emission-line gas. The combination of the
harder quasar radiation field, powerful jets and the larger amount of
gas accounts for their larger ELR sizes and luminosities of HEGs than
LEGs.

However, is the AGN continuum (accretion disk and jet) really enough
to account for the emission line properties we observe?  Different
specific aspects of the ionized gas suggest that other sources of
high-energy photons may contribute to the photoionization of the gas.

The presence of tails, bridges, shells, irregular features and
amorphous halos observed in the emission-line maps of some HEGs
(3C~33, 3C~180, and 3C234) are suggestive of traversal jet interaction
with the ISM. The outward motion of the radio jet may result in the
passage of shocks through the clouds. The shock occurs at the
interface between the outflowing plasma and the ambient medium, such
as in the hot spots and on the lateral edges of the jet cocoon, a
crucial site of gas photoionization. \citet{couto17} observed
increased line ratios for 3C~33 outwards, using Integral field
spectroscopy with Gemini, and interpreted them as due to a lateral
expansion of the ambient gas in the nuclear strip due to shocks
produced by the pass age of the radio jet. Supernovae and past radio
outbursts can also produce multiple shocks through the ISM and cause
line-emitting regions cospatial with the radio emission (e.g.,
\citealt{rampadarath18}).  In the case of our emission maps, the
region of multiple shocks is clear for 3C~180, where at the edges of
the radio lobes, the gas is pressured and ionized. These emission
features are more common for FR~IIs because they have larger jet bulk
speeds and larger radio lobes than FR~Is. This difference would
contribute to explain why FR~IIs (and then HEGs) show more extended
and variegated morphologies of ionized gas than FR~Is.

Another source of high-energy photons are the young, hot
stars. \citet{Robinson87} concluded that the very extended
emission-line gas in RGs can be photoionized by typical O and B
stars. The extended disturbed patches of ionized gas observed in
3C~33, 3C~234 and 3C~285 seem to be associated with regions of intense
star formation, induced by jet shocks or by a tidal
merger/interaction, as suggested by the HST images
\citep{madrid06,floyd10,baldi08}. These sources are all classified as
HEGs.  Indeed, this ionization mechanism may contribute more for the
HEGs which show evidence of ongoing star formation rather than more
passive host galaxies of LEGs (e.g.,
\citealt{baldi08,baldi10a,herbert10,janssen12,hardcastle13,gurkan15,westhues16}). Post-AGB
stars which also emit in UV can also play a minor role in the
photoionization.

\subsection{Covering factor}
\label{coverdiscuss}

The mean covering factor we measured from the HST images for our
objects, 0.66, is in agreement with the broad range of values derived
from previous multi-band studies for radio-quiet and -loud AGN in the
same interval of nuclear luminosities (0.1--0.8, e.g.
\citealt{Netzer93,wills93,maiolino95,Zheng97,Laor97,Maiolino01,capetti05b}). This
result also reconciles with the relative weakness of the NLR with
respect to the BLR, typically 0.1-0.01 per cent
\citep{boroson92,osterbrock82,osterbrock89,buttiglione09}.

We also estimated covering factors larger than the unity for four
sources, belonging to both LEG and HEG populations. This result
indicates that the line emission measured from the HST images is
larger than what we expected from the ionizing photon
budget. Different aspects might cause an overestimation of the the
covering factors. A systematic larger error of the line flux density
for a partially incorrect continuum subtraction in the bandwidth can
affect our $\Omega_{\rm NLR}$ estimates. The underestimation of the
power of other contributing ionizing sources, i.e. jet shocks and star
formation in the ISM in the outer region of the ELR, can lead to a low
photon ionizing budget. A small change in the X-ray spectral index or
the bolometric correction factors, which are known to vary broadly for
different AGN classes \citep{ho08,lusso12}, could lead to a
significant change in the estimation of the photon budget Q$_{\rm
  ion}$. The assumption of the isotropy of the ionizing nuclear source
might have an effect on the covering factors. Marginal effect on the
$\Omega_{\rm NLR}$ could come from relativistic beaming in optical
nuclei \citep{capetti05b} and X-ray nuclear variability.

Overall, our study shows that HEGs have slightly larger covering
factors than LEGs, despite consistent within the errors. In line with
our results we found for the two populations, this $\Omega_{\rm ELR}$
difference can be interpreted as due to the the combined effect of the
stronger central engine and larger availability of cold gas, valid for
HEGs.

\subsection{[O~III]/H$\alpha$ across the galaxy and feedback}

The emission-line ratios depend on both the gas density and the gas
proximity to the ionizing source
\citep[e.g][]{Jackson97,kewley06}. Therefore, one would expect
low-excitation lines such as H$\alpha$ to be present at larger scales
in the galaxy with respect to high-excitation lines, such as
[O~III]. However, for at least 5 ($\sim$26\%) galaxies in our sample,
all HEGs (3C~33, 3C~180, 3C~227, 3C~234, 3C~285), [O~III] extended
emission is clearly observed at larger scales (on kpc scale) than that
of H$\alpha$. These results are similar to previous studies on ELR in
radio-quiet and -loud galaxies
\citep{Robinson87,Storchi-Bergmann96,capetti96,Robinson00,Taylor03,kraemer08,maksym16}. The
observed line ratios are consistent with the idea of a mixed medium,
with a variegated density and optical depth even at large distance
from the central BH illuminated by various ionizing sources along the
optical axis \citep[e.g.][]{Binette96}.

Furthermore, larger amplitude variations of the [O~III]/H$\alpha$
ratios have been observed across the extension of the ELR for HEGs,
rather than for LEGs (Fig.~8).  Therefore, in light of our results,
how can we explain this high [O~III]/H$\alpha$ at large distance from
the ionizing nucleus and large variation of this ratio?  Different
scenarios might account for the observed ratios.

It is possible that the lack of extended H$\alpha$
emission compared to that of [O~III] is related to the sensitivity of
our {\it HST} observations. For example, in the case of 3C~227, our
images show faint extended [O~III] emission (over 20 kpc) towards the
southeast of the galaxy that is not visible in the corresponding
H$\alpha$ image. However, the same structures are clearly
visible in both the [O~III] and H$\alpha$ images of the galaxy
studied by \cite{Prieto93}.

Another possibility is an inadequate correction of
[N~II]$\lambda$$\lambda$6549,6583\AA~emission lines to the flux within
the bandpass of filter used to image the H$\alpha$ emission. This is
due to the fact that [N~II] line may change cross the spatial extent
of the galaxies. In fact, long-slit spectroscopic studies along the
NLR reveal line ratio H$\alpha$/[N~II]$\lambda\lambda$6949,6983
changes by a factor of 1.0 - 5.3 from the nuclear to the extended
regions of the galaxies
\cite[e.g.][]{Robinson87,Storchi-Bergmann96,Robinson00,Taylor03}. Therefore,
assuming that a similar change occurs in our maps, this effect would
overestimate the [O~III]/H$\alpha$ line ratios by a factor of $\sim$2
at large scales. This effect might be important for galaxies with
substantial extended emission, such as HEGs.

Another factor which might affect the [O~III]/H$\alpha$ ratio is the
reddening. Due to dust mostly in the central regions, the H$\alpha$
emission at large distance should be less affected than the [O~III]
emission. This is the case for 3C~33, 3C~136, and 3C~285 which show
dust features across the optical galaxy and decoupled [O~III] and
H$\alpha$ emission in our maps.

Assuming the observed line ratio scatters are real, physical scenarios
can be discussed in the framework of feedback processes. Both
anisotropic AGN illumination and jet-cloud interaction, which produce
radiation continuum over a wide range in ionization parameters, can
account for the ionization states of both the nuclear and the
off-nuclear emission line gas \citep{Robinson87,tadhunter02b}. In the
context of a quasar-mode feedback, theoretically, a standard-disk
ionizing continuum, which probably includes combination of power-laws,
thermal bumps, a high-energy tail and outflowing gas
\citep{binette88}, predicts the observed strengths of high-ionization
lines, mainly [O~III] \citep{Robinson87}. This ionizing source can
also explain the large [O~III]/H$\alpha$ observed across the galaxy in
HEGs. The kinetic energy released by the jets and shocks are important
sources of ionizing photons in RGs, for both the AGN classes. The
local kinematics and the gas pressure, in an event of gas expansion
due to the jet passage, should be considered since they surely play a
role in the photoionization mechanism (see the integral field spectra
of 3C~33 \citealt{couto17}). Shocks can can mimic the emission line
ratios typical of a LEG \citep{allen08,best00a}. Furthermore,
different sources of high-energy photons located across the galaxy
extension, might provide further ionization at large distances from
the center. Sparse knots of star formation and post-AGB stars, can
produce a substantial diffuse field of ionizing photons: the resulting
emission line ratios cover the whole region typical of LEGs and HEGs
in the diagnostic diagrams \citep{binette94,stasinska08,sarzi10}. In
conclusions, all the various sources of energetic photons present
across the galaxy have an important impact on the multi-phase gas of
the galaxy (positive and negative feedback,
\citealt{morganti17,cresci18}) and contribute to the large scatter
observed in [O~III]/H$\alpha$ maps.

The combination of the primary central quasar with the jets and the
secondary ionizing sources located along the optical axis, would
explain the large emission-line ratio variation observed in our maps
of the HEG population. Conversely, the lack of star formation and of a
bright quasar makes the emission line morphologies of LEGs more
homogeneous and neat. The LEG population of our sample fits in a more
simple picture where the synchrotron photons mainly ionize the
surrounding gas and account for the small scatter observed in the
[O~III]/H$\alpha$ ratios.  Old stellar population, tracing the smooth
elliptical galaxy, might marginally contribute to ionize the gas in
the central region of the host.

\section{Summary and conclusions }

We have presented HST/ACS narrow-band [O~III] and H$\alpha$ images for
a sample of 19 RGs with redshifts $<$0.3. We studied the morphology of
the emission-line regions and the ratio between the two lines. Based
on the emission-line ratios from nuclear long-slit spectra, we divided
the sample into HEGs and LEGs, which are known to have two different
accretion modes: cold-gas and hot-gas accretors, respectively. The
line properties and the linear correlation between radio core power,
X-ray luminosities and the total emission line luminosities suggest
two different pictures for the two classes:

\begin{itemize}

\item LEGs show more compact ELR on scale of kpc but smaller than
  HEGs. The mean covering factor is 0.60 associated with ionized
  masses of 4.7 $\times$ 10$^5$ M$_{\odot}$. The variation of the
  [O~III]/H$\alpha$ ratio across the ELR is smaller than the sharp
  changes observed for HEGs. All these properties of ELR reconcile
  with a picture of a LEG, powered by a synchrotron-dominated
  nucleus. In fact, the main source of high-energy photons is the
  radio jet, which ionize a secular gas-poor environment.

\item HEGs produce more prominent ELR properties than LEGs. Their
  structures are more extended, on scale of some kpc, and more
  disturbed than LEGs line regions. The covering factors are slightly
  larger than LEGs and the involved ionized masses are larger, up to
  10$^{7.70}$ M$_{\odot}$. The [O~III] gas is observed also at large
  distances from the nucleus, making the [O~III]/H$\alpha$ ratios
  change rapidly across the ELR. All these properties of their ELR
  match with the picture of HEGs, powered by a quasar (standard
  accretion disk), which contributes to illuminate the gas-rich ISM
  together with the radio jets.  However, to justify the emission-line
  ratio changes and the morphologies, we need to invoke further
  photoionizing sources.  Star-formation and shocks from the radio
  jets contribute to the photoionization of the clouds along the
  radio/optical axis.
\end{itemize}

Based on our current comprehension of the radio AGN phenomenon
(e.g. \citealt{best12,heckman14}), we can put our results into the
following galaxy-AGN evolutionary scenario: a link between the
accretion and the photoionization modes valid for radio-loud AGN is
emerging.  On one hand, LEGs, which live in rich environments
\citep{hardcastle04,tasse08,ramos-almeida13,gendre13,castignani14,ineson15},
have gone through a 'dry' merger process \citep{balmaverde06core} in
the past, lacking dust and a cold gas reservoir. The paucity of such a
gas prevents the formation of a radiatively efficient accretion
disk. The radio jet, sustained by the hot-phase accretion, is the main
source of ionizing photons, sufficient to account for the compact weak
line regions observed in LEGs. On the other hand, HEGs have been
recently engaged in galaxy-galaxy interaction/mergers ('wet' merger)
\citep{baldi08,Ramos-Almeida11,Ramos-Almeida12,herbert10,hardcastle13,chiaberge15},
which bring new fresh cold gas and can trigger either star formation
and/or cold-gas accretion. This cold gas reservoir feeds the standard
accretion disk, which in turn supports the launch of powerful
jets. These ionizing sources mainly illuminate the NLR, creating
bright extended disturbed emission line morphologies, as observed in
our HST maps. Their line morphologies also resemble their optical-UV
galaxy disturbances \citep{baldi08}, suggestive of an ISM which has
not settled down yet after the merger.  Star formation across the
galaxy and shocks might contribute to the photoionization at large
distances from the nucleus.

The positive feedback of the jet and the central AGN on the ISM are
also supported by the alignment of the radio axis with the morphology
of the ELR. While what expected from a radio-mode feedback matches the
emission-line properties of the LEGs, the quasar-mode feedback plays a
crucial role to interpret the ELR properties of HEGs, similar to what
observed in local radio-quiet Seyferts/QSO
\citep{williams17,fisher18,revalski18}. In both the feedback modes,
the gas is photoionized along the axis of the jet, where most of the
ionizing photons are able to escape and most of the cold gas is
shocked. Hydrodynamic models of jet expansion in a two-phase ISM,
which predict double or multiple bubbles with velocity fields ranging
between 500 and 1000 km s$^{-1}$, indicate that the kinetic energy of
this outflow transfers from 10\% to 40\% of the jet energy to the cold
and warm gas \citep{wagner12}, resulting in gas ionization along the
radio jet. A radio-mode feedback is equally important for LEGs and
HEGs \citep{whittam18}. However the AGN illumination needs to be also
taken into account to interpret the ELR properties of HEGs. The
interplay between bulk flow and turbulence from a jet and the
bi-conical illumination from a quasar, are important aspects to stir up
and photoionize the ISM, predicting broadly the disturbed
morphologies and properties of the ELR for HEGs.

We estimated a mean emission-line covering factors of 0.66 for our
sample. The marginally lager covering factors found for HEGs than LEGs
is probably the result of the different conditions present in the
hosts of two classes. In the case of LEGs, the paucity of cold gas and
the weaker radiation field produced by the jet result in smaller
covering factors. As observed in nearby Seyferts
(e.g. \citealt{contini12,villar15,freitas18}), a radiation-bound
photoionization model from the AGN, outflow, jet shocks and young
stars and the large amount of gas clouds support the extreme line
properties observed in HEGs and the larger covering factors
\citep{hainline14,dempsey18}.

The advent of the high-resolution integral field era, i.e. VLT/SINFONI
and MUSE, is bringing interesting results on the study of jet--ISM
interplay in local galaxies (e.g.,
\citealt{santoro15,dasyra015,cresci15,salome16,roche16,balmaverde18}). However,
the AGN feedback is still an appealing and complex phenomenon, on
which this new era of telescopes will cast more light to understand
this mechanism in the next future.

\acknowledgments

The authors wish to thank A. Laor for the insightful discussion,
E. Torresi for providing the X-ray spectral analysis of the sample,
and D.~R.~A. Williams for the helpful comments on the line properties
of the sample.  We also thank the reviewer for the useful comments,
which helped us improve the quality of the manuscript. RDB and IMC
acknowledge the support of STFC under grant [ST/M001326/1].

\appendix

\section{A.1. Control check of the [O~III]/H$\alpha$ spatial distribution}
\label{check}

\begin{figure*}
\begin{tabular}{cc}
\hspace*{0.5cm}\includegraphics[scale=0.9]{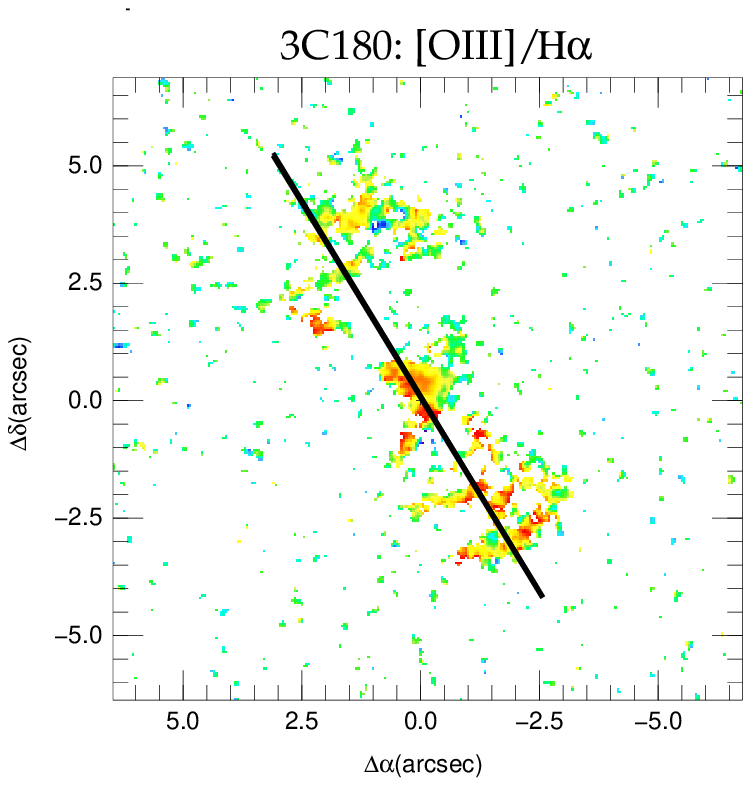}&
\hspace*{0.5cm}\includegraphics[scale=0.9]{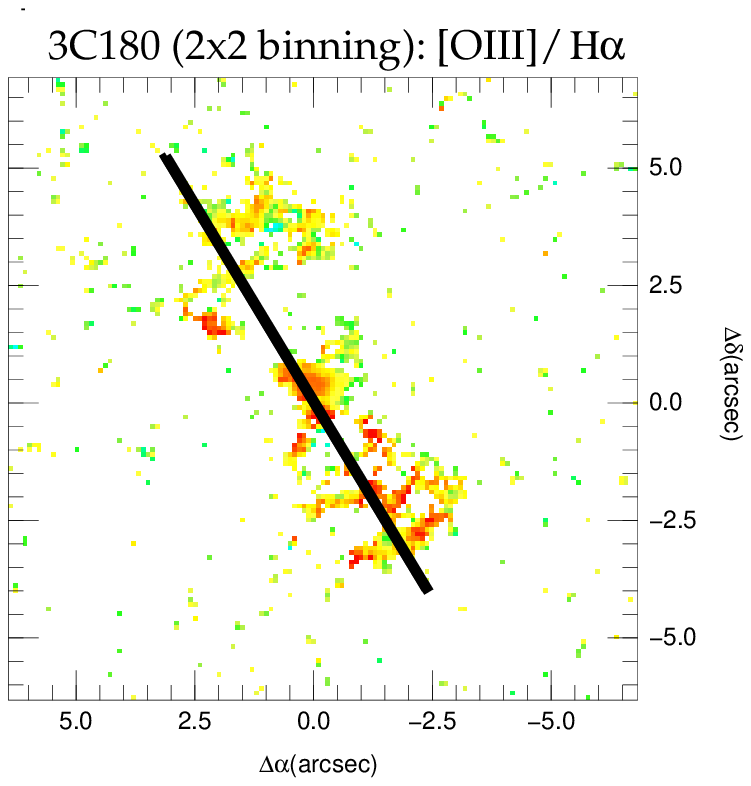}\\
\hspace*{0.1cm}\includegraphics[scale=0.4]{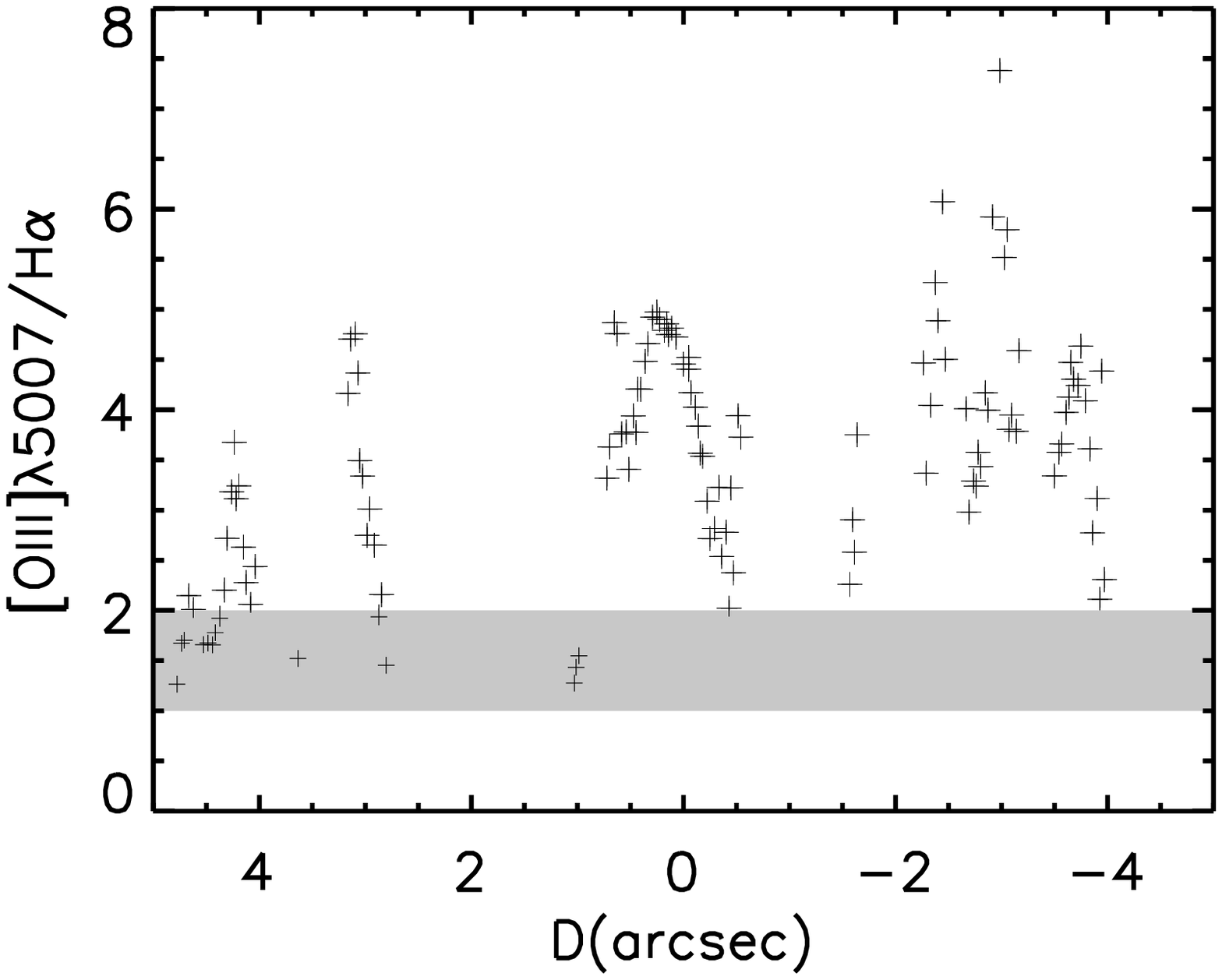}&
\hspace*{-0.7cm}\includegraphics[scale=0.4]{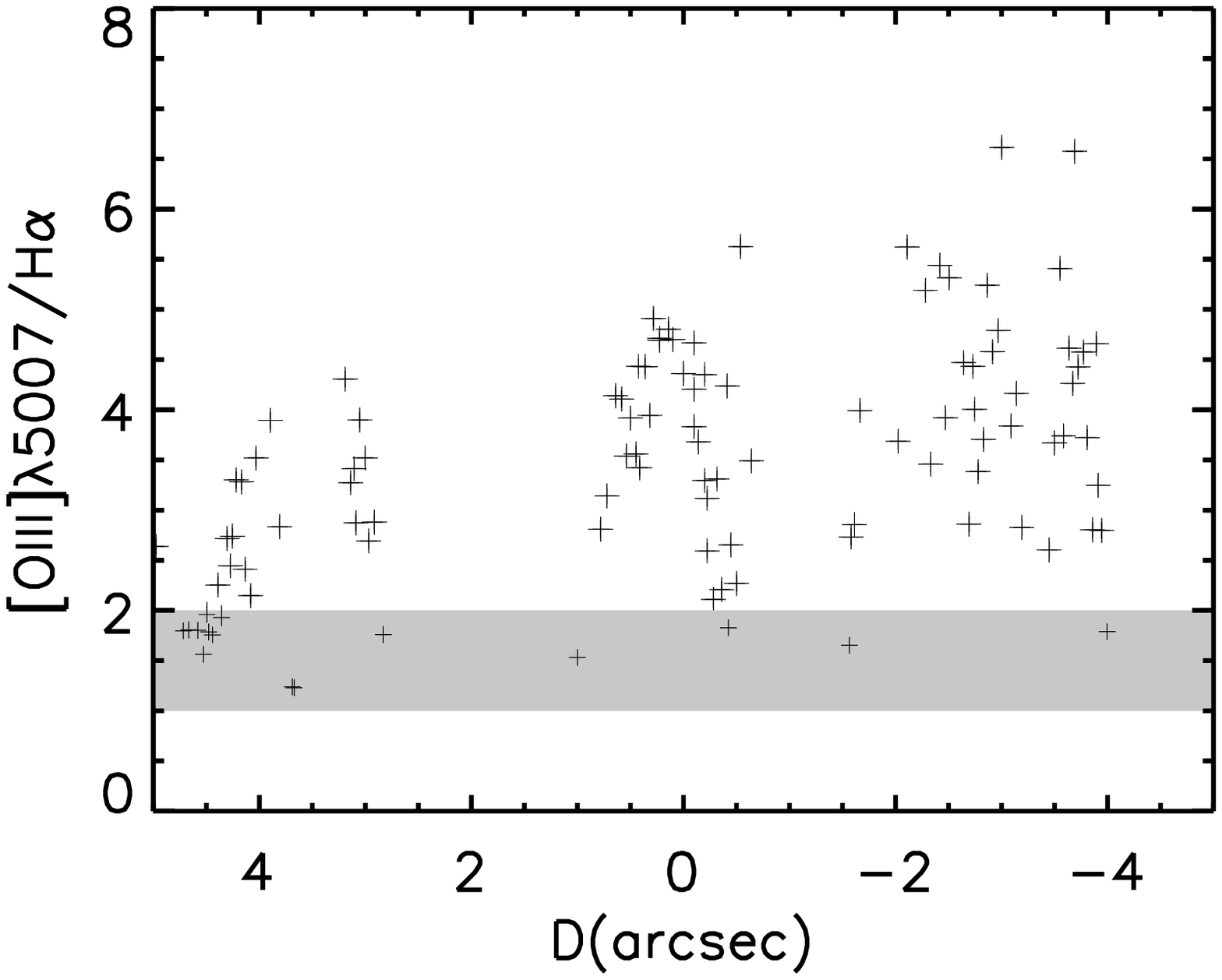}\\
\end{tabular}
\caption{Top left: the spatial distribution of the [O~III]/H$\alpha$
  line ratio for 3C~180. Top right: the spatial distribution of the
  [O~III]/H$\alpha$ line ratio obtained after binning the [O~III] and
  the H$\alpha$+N~II images using 2 $\times$ 2 pixel binning mode. The
  black-solid lines drawn in the figures correspond to a section of
  the galaxy of 3 pixels of width, extended along the optical axis of
  the source. Bottom left: the [O~III]/H$\alpha$ line ratios plotted
  against the distance (D) from the center of the galaxy. Bottom
  right: same as bottom left but for the top right figure. When
  generating these plots we used the section of the galaxy
  corresponding to the black, solid line in the top-left panel. In
  addition, we used the values in Column 6 and 7 in
  Table~\ref{detect2} to correct for [O~III]$\lambda$4959 and
  [N~II]$\lambda\lambda$6949,6983 contribution to the flux in the
  band. As clear from the figure, the emission-line structures
  observed in the unbinned image are still visible after the binning.}
\label{3C180_binning}
\end{figure*}

To test the potential effects of misalignment and/or low
signal-to-noise of the two emission-line ratio maps, we first shift
the narrow band images prior to the continuum subtraction using 0.5
and 1 pixel shifts in each direction, x and y. Secondly, the [O~III]
and the H$\alpha$ images are binned using a 2 $\times$ 2 pixel binning
mode with the aim of increasing the S/N and reduce the effects of any
small mismatch. An example of the `$2 \times 2$ binning' test is shown
in Figure~\ref{3C180_binning} using 3C~180. The plots corresponding to
each galaxy shown in Figure~\ref{ratiovsdistance} are generated after
each of these tests. Although the individual values of each pixel
changed, the general trends remain the same.

The [O~III]/H$\alpha$ images of the galaxies are systematically
observed using the same chip and the same pointing set up, within the
same orbit and using the same guiding star. Therefore, if the
contribution of the continuum emission in the HST band is not
significantly important, one could directly divide the two images
product of the {\it HST}/ACS pipeline before any further reduction. If
registration-image issues are not important, the same structures
observed in the `processed images' should still be visible. We perform
this test for the particular cases of 3C~33 and 3C~234, two of the
most spectacular objects in our sample. The morphologies and the shape
of the scatter are similar to those observed after the full reduction
process.

Overall, we conclude that the the [O~III]/H$\alpha$ variations and the
scatter seen in Figures~\ref{panel} and \ref{ratiovsdistance}
respectively are generally reliable, and trace the spatial variations
of the gas with different optical thickness and/or densities along the
optical axis.

\bibliography{my}

\end{document}